# Tevatron Electron Lenses: Design and Operation

*Vladimir Shiltsev[1*], Kip Bishofberger[2], Vsevolod Kamerdzhiev[1], Sergei Kozub[5], Matthew Kufer[1], Gennady Kuznetsov[1], Alexander Martinez[1], Marvin Olson[1], Howard Pfeffer[1], Greg Saewert[1], Vic Scarpine[1], Andrey Seryi[3], Nikolai Solyak[1], Veniamin Sytnik[5], Mikhail Tiunov[4], Leonid Tkachenko[5], David Wildman[1], Daniel Wolff[1], and Xiao-Long Zhang[1]*

*[1]Fermi National Accelerator Laboratory, PO Box 500, Batavia, IL 60510, USA*

*[2]Los Alamos National Laboratory, Los Alamos, NM 87545, USA*

*[3]Stanford Linear Accelerator Center, Stanford, CA 94025 , USA*

*[4]Budker Institute of Nuclear Physics, Novosibirsk, 630090, Russia*

*[5]Institute of High Energy Physics, Protvino, 142284, Russia*

[*] e-mail: shiltsev@fnal.gov

## Abstract

The beam-beam effects have been the dominating sources of beam loss and lifetime limitations in the Tevatron proton-antiproton collider [1]. Electron lenses were originally proposed for compensation of electromagnetic long-range and head-on beam-beam interactions of proton and antiproton beams [2]. Results of successful employment of two electron lenses built and installed in the Tevatron are reported in [3,4,5]. In this paper we present design features of the Tevatron electron lenses (TELs), discuss the generation of electron beams, describe different modes of operation and outline the technical parameters of various subsystems.

## 1. Introduction

Fermilab's Tevatron is currently the world's highest energy accelerator in which tightly focused beams of 980 GeV protons and antiprotons collide at two dedicated interaction points (IPs). Both beams share the same beam pipe and magnet aperture and, in order to avoid multiple detrimental head-on collisions, the beams are placed on separated orbits everywhere except the main IPs by using high-voltage (HV) electrostatic separators. The electromagnetic beam-beam interaction at the main IPs together with the long-range interactions between separated beams adversely affect the collider performance, reducing the luminosity integral per store (period of continuous collisions) by 10-30%. Tuning the collider operation for optimal performance becomes more and more cumbersome as the beam intensities and luminosity increase. The long-range effects which (besides being nonlinear) vary from bunch to bunch are particularly hard to mitigate. A comprehensive review of the beam-beam effects in the Tevatron Collider Run II can be found in Ref. [1].

Electron lenses were proposed for compensation of the beam-beam effects in the Tevatron [2]. An electron lens employs space-charge forces of a low-energy beam of electrons that collides with the high-energy bunches over an extended length $L_e$. The lens can be used for linear and nonlinear beam-beam force compensation depending on electron current-density distribution $j_e(r)$ and the ratio of the electron beam radius $a_e$ to the rms size $\sigma$ of the high-energy beam at the location of the lens. The electron current profile (and thus the radial dependence of electric and magnetic forces due to electron space charge) can be easily changed for different applications. The electron beam current can be adjusted between individual bunches, equalizing the bunch-to-bunch differences and optimizing the performance of all bunches in a multi-bunch collider.

A shift of the betatron frequency (tune) of high-energy particles due to EM interaction with electrons is a commonly used "figure of merit" for an electron lens. A perfectly steered round electron beam with current density distribution $j_e(r)$, will shift the betatron tunes $Q_{x,y}$ of high-energy (anti-)protons by [2]:

$$dQ_{x,y} = \pm \frac{\beta_{x,y} L_e r_p}{2\gamma e c} \cdot j_e \cdot \left( \frac{1 \mp \beta_e}{\beta_e} \right) \qquad (1),$$

where the sign reflects focusing for protons and defocusing for antiprotons, $\beta_e = v_e/c$ is the electron beam velocity, $\beta_{x,y}$ are the beta-functions at the location of the lens, $L_e$ denotes the effective interaction length between the electron beam and the protons or antiprotons, $r_p = e^2/mc^2 = 1.53 \cdot 10^{-18}$ m is the classical proton radius, and $\gamma_p = 1044$ the relativistic Lorentz factor for 980-GeV protons. If the electron beam is wider than the (anti)proton beam, then all of the high-energy particles acquire the same $dQ_{x,y}$. The factor $1 \pm \beta_e$ reflects the fact that the contribution of the magnetic force is $\beta_e$ times the electric force contribution, and its sign depends on the direction of the electron beam (in this case, both Tevatron Electron Lenses (TELs) direct the beam against the antiproton flow, which corresponds to $1+\beta_e$ and maximizes the tuneshift). Both TELs operate with only a few Amperes of electron current at up to 10keV electron energy and can shift the betatron tune by as much as $dQ_{x,y}^{max} \approx 0.008$ [4].

Two Tevatron Electron Lenses (TELs) were built and installed in two different locations of the Tevatron ring, A11 and F48. Figure 1 depicts the general layout of the TEL-1 and TEL-2. The electron beam is generated by a thermionic gun immersed in the solenoidal magnetic field. Strongly magnetized electrons are accelerated to the kinetic energy of 5-10 kV and follow the magnetic field lines into the main superconducting solenoid where the interaction with high energy proton/antiproton bunches occures. While the high energy particles continue on (along) the Tevatron orbit, the low energy electrons exiting the main solenoid are guided into the collector and are not being recirculated (reused). A list of the TEL and relevant Tevatron parameters is given in Table 1. Both lenses were successfully tested and used in two regimes of operation – a) for compensation of beam-beam effects [3, 4] and b) for removal of uncaptured particles from the abort gaps between the bunch trains [5]. Three conditions were found to be crucial for successful operation of the electrons lenses [4]: 1) the electron beam must be transversely centered on the proton (antiproton) bunches, within 0.2–0.5 mm, along the entire interaction length; b) fluctuations in the electron current need to be less than one percent, and the timing jitter within one nanosecond, in order to minimize emittance growth of the high-energy beams; and c) the transverse profile of the current density should have specific shape, depending of application, e.g. a distribution featuring a flat-top and smooth edges is needed for compensation of long-range beam-beam effects.

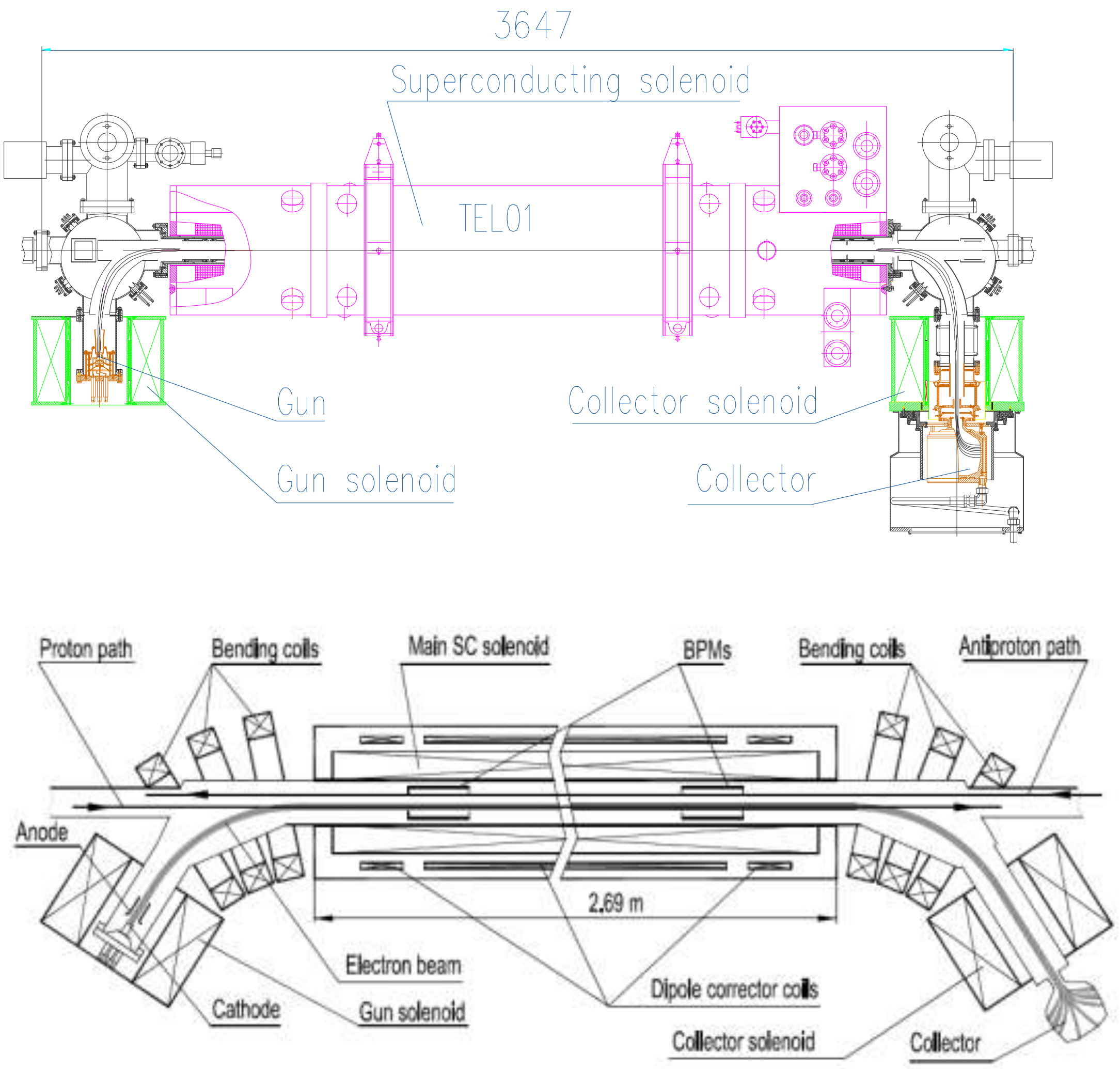

Figure 1. (top) General layout of the TEL-1 installed at F48 location, top view.

(bottom) General layout of the TEL-2 installed at A11 location, top view.

In this paper, we present main design features of the Tevatron Electron Lenses, describe in detail major subsystems and discuss the experience with the lenses in the Tevatron.

# 2. Magnetic and cryogenic systems

The main requirements for the TEL magnetic system have been formulated in Ref. [2]. Electron beam needs to be strongly magnetized at every point of its path from cathode to collector in order to: a) keep the beam stable against its own space charge forces, b) not be disturbed by electromagnetic forces of (anti)proton bunches passing though it, and c) be rigid enough and not cause any coherent instabilities in the Tevatron high-energy beams. Besides being used for the transport of electrons from the cathode to the collector, the magnetic system must be capable of changing – by adiabatic magnetic compression - the electron beam size in the interaction region, and allow precise positioning of the electron beam on the high-energy beam of choice. The three solenoids in the TEL-1 are oriented as shown in Fig.1. The gun solenoid sits in the lower-left corner perpendicular to the long Tevatron beam pipe, the main solenoid surrounds the beam pipe, and the collector solenoid resides in the lower right. The geometrical center of the Tevatron vacuum chamber is precisely aligned with the magnetic axis (center) of the main solenoid Electrons, originating from the electron gun, follow the magnetic field lines, bent in the horizontal plane. The solenoids were manufactured at the Institute of High Energy Physics in Protvino, Russia and tested at Fermilab. Many technical details on the magnet construction and magnetic field simulations can be found in Refs. [6,7].

## *2.1 Main SC and conventional solenoids*

The transverse cross section of the main TEL solenoid is shown in Fig.2. It is capable of reaching the maximum field of 6.5 T at 1780 A and liquid Helium temperature of 4.6-5.3 K. The main solenoid does not contain a closed current loop; when energized, the current flows out of its current leads and through external power supplies. The main solenoid uses NbTi wire intertwined with copper wire (Cu/ NbTi ratio of 1.38), rated for 550A at a temperature of 4.2 K; the wire itself measures 1.44mm by 4.64mm cross-section. The cable is wrapped by polyimide film of 0.03 mm thickness with 1/3 overlap. The main SC coil is wound on stainless steel tube of 151.4 mm diameter and 4 mm thickness. The frame insulation is three layers of polyimide film of 0.1 mm thickness. A 4.85cm-thick, low-carbon steel shield wraps around the coils, which enhances the field strength, keeps the field lines compressed near the solenoid's ends, improves the homogeneity throughout the interaction region, and reduces stray fields. While the solenoid is designed to handle 6.5T, its nominal operating strength is 3.0-3.5T. During initial operation with the main solenoid, it successfully reached 6.7T before quenching.

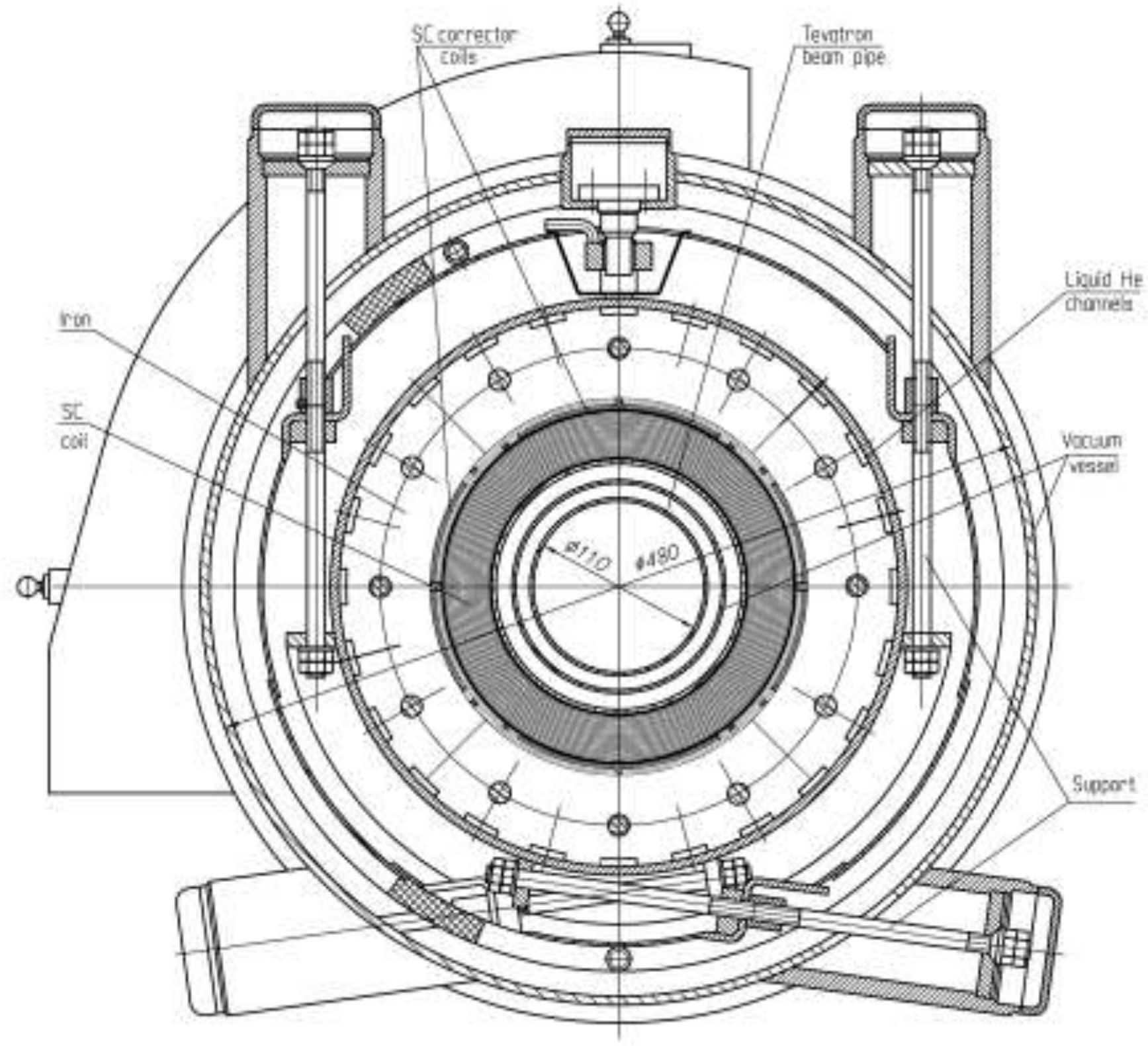

Figure 2. Transverse cross-section of the main SC solenoid.

The gun and collector solenoids use water-cooled copper windings which generate maximum of about 0.4T field on axis with maximum 340A of current. The resistance and inductance of the 391 turns of wire is roughly 0.19Ω and 18mH. The bore of each magnet has a diameter of about 24.0cm and a length of 30.0cm, enough to contain the electron gun and the collector input port. There is a small design difference between the gun and collector solenoids - the collector solenoid has an additional iron plate on its back end, which reduces the field strength outside the solenoid (in the region of the collector itself). Electron beam shape and position correctors are set inside each of the conventional solenoids. The corrector consists of four coils, which can be commutated either as a quadrupole or as two dipoles (vertical and horizontal). Each coil has layer shape geometry with $0.74^{\circ}$ inner and $40.04^{\circ}$ outer angles, 11.2-cm inner radius and 0.9-cm thickness. The length of coil is equal to 30 cm. The coils were wound by 1-mm diameter copper wire and have 620 turns each. In dipole configuration the field is equal to 19 G/A; the quadrupole field is equal to 6 G/cm/A. Maximum current in these coils does not exceed 5A. In routine operation, we rarely employed these corrector coils in the gun and collector solenoids.

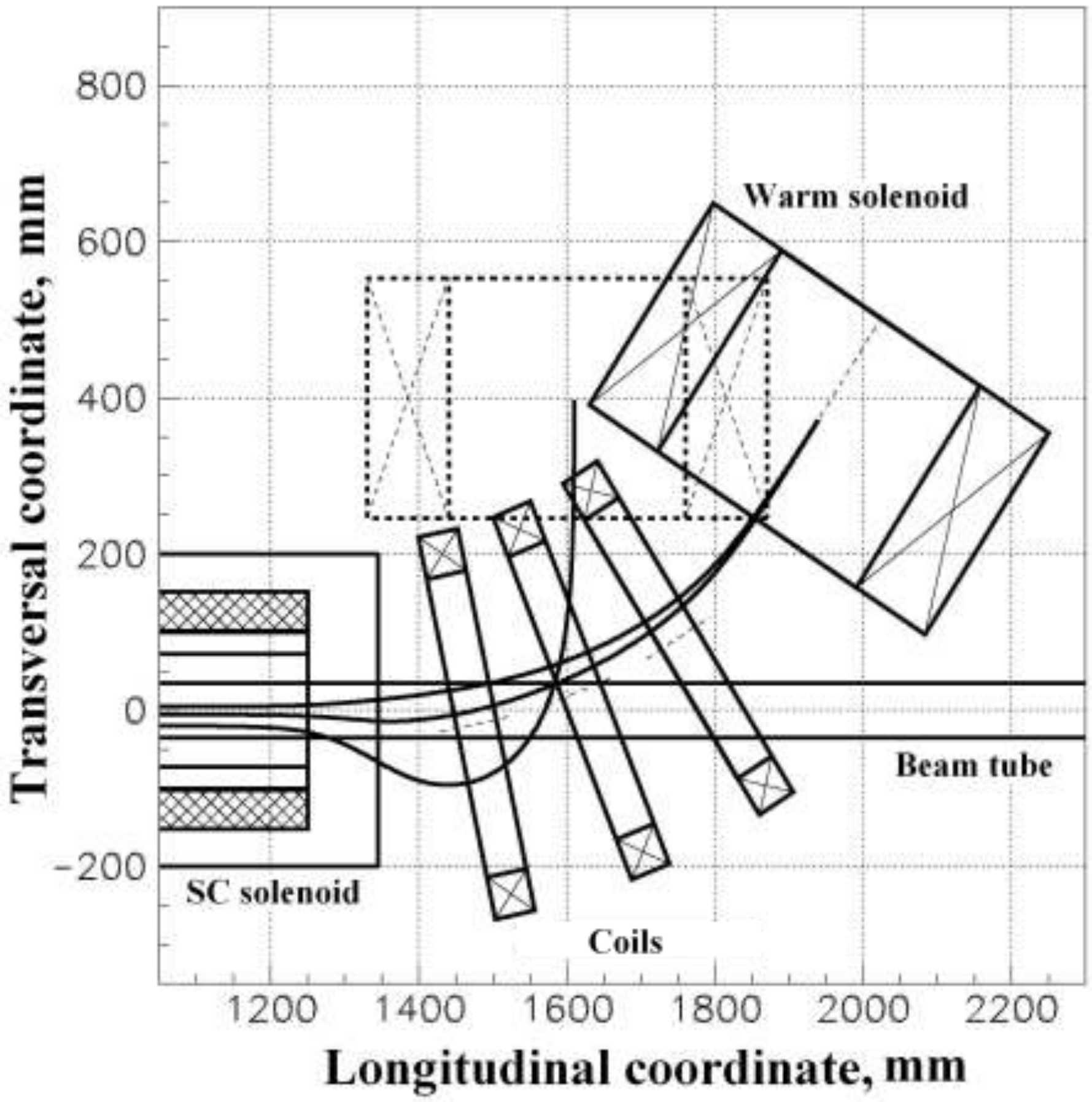

Figure 3. Simulations of the magnetic field lines in the bending sections of the TEL-1 and TEL-2 carried out using the MULTIC code [8]. The placement of the TEL-1 gun solenoid is shown by dashed lines; TEL-2 magnets are represented by solid lines. Magnetic field lines (electron beam trajectories) in both TELs are shown as well.

Axes of the gun and collector solenoids in TEL-1 are perpendicular to the axis of main SC solenoid. Operational experience with such a configuration has shown that the electron beam transmission can be assured only within the limited range of the main solenoid field to gun and collector solenoid field ratio $B_{main}/B_{gun} \approx 10\text{-}20$ [9]. Beyond this range, the electron beam did not fit the aperture of the electrodes in the bending section of the TEL-1. In addition, there was a significant – several mm - vertical drift of the electron beam due to $\boldsymbol{B}\times\nabla\boldsymbol{B}$ effect in the bending section, which scales as :

$$dy(z) = \int_s \frac{2U_e}{e\beta_e B(z)R(z)}dz \qquad (2),$$

where $z$ is the coordinate along the electron trajectory, $U_e$ is the electron beam kinetic energy, $B(z)$ and $R(z)$ are the magnetic field and the magnetic field line curvature radius. The second electron lens (TEL-2) was designed to significantly increase both $B(z)$ and $R(z)$ in the bending sections, reduce the drift $dy(z)$ 4-5 fold and allow wider range of the ratios $B_{main}/B_{gun}$ . For that, axes of the gun and collector solenoids (identical to those in TEL-1) were set at 57° with respect to the main solenoid axis and three additional short solenoids were added in each bending section, as shown in Fig.3. Each of the three new coils generated about 420 G of magnetic field in its center. All the coils were powered in series with the gun or collector solenoids. As the result, the minimum magnetic field in the bending section has been increased from 800 G to 1300-1800 G for the typical field configuration . The electron beam size in the bending region is reduced as well, as it scales as $a_e(s)=a_{cathode}(B_{main}/B(s))^{1/2}$. Therefore, the ratio of the gun solenoid field to the main solenoid field can now be varied in a much wider range allowing a greater adjustment flexibility of the electron beam size in the interaction region. For example, for $B_{gun} = B_{collector}$ =0.3T the electron beam can pass the main solenoid with $B_{main}$= 0.3-6.5T in the TEL-2 while in the TEL-1 the allowed main solenoid field range was limited from 2.7T to 5.5T.

## *2.2 Corrector magnets*

The TEL’s ability to adjust electron trajectory inside the main solenoid to the straight line of (anti)proton orbit is needed in four degrees of freedom: the upstream position and the angle, both in the horizontal and vertical directions. Six superconducting dipole corrector magnets are used for this steering. Two of these correctors, one oriented horizontally and one vertically, are located at the upstream end of the main solenoid; their goal is to adjust the upstream transverse position of the electron beam to equal that of the (anti)proton orbit. Two other correctors extend nearly the length of the main solenoid. These long correctors have the ability to angle the electron beam along their entire length. Once the upstream correctors are set, the long correctors are adjusted so that the electron beam coincides with the (anti)proton orbit, as drawn in Fig.4. Beam position monitors (described in the next section) situated at the upstream and downstream ends of the long correctors are used to confirm that the two species (electrons and antiprotons or electrons and protons) are set at identical transverse positions. The electron beam can end at a variety of positions, yet it must be able to pass into the collector. To accomplish this, a third set of correctors are located downstream of the long correctors in order to steer the beam back into a position where it will successfully enter the collector. These correctors, identical to the upstream correctors, often are adjusted simultaneously with either the upstream or the long correctors, but in the opposite direction; in this sense, they ``undo" the changes made by the other correctors.

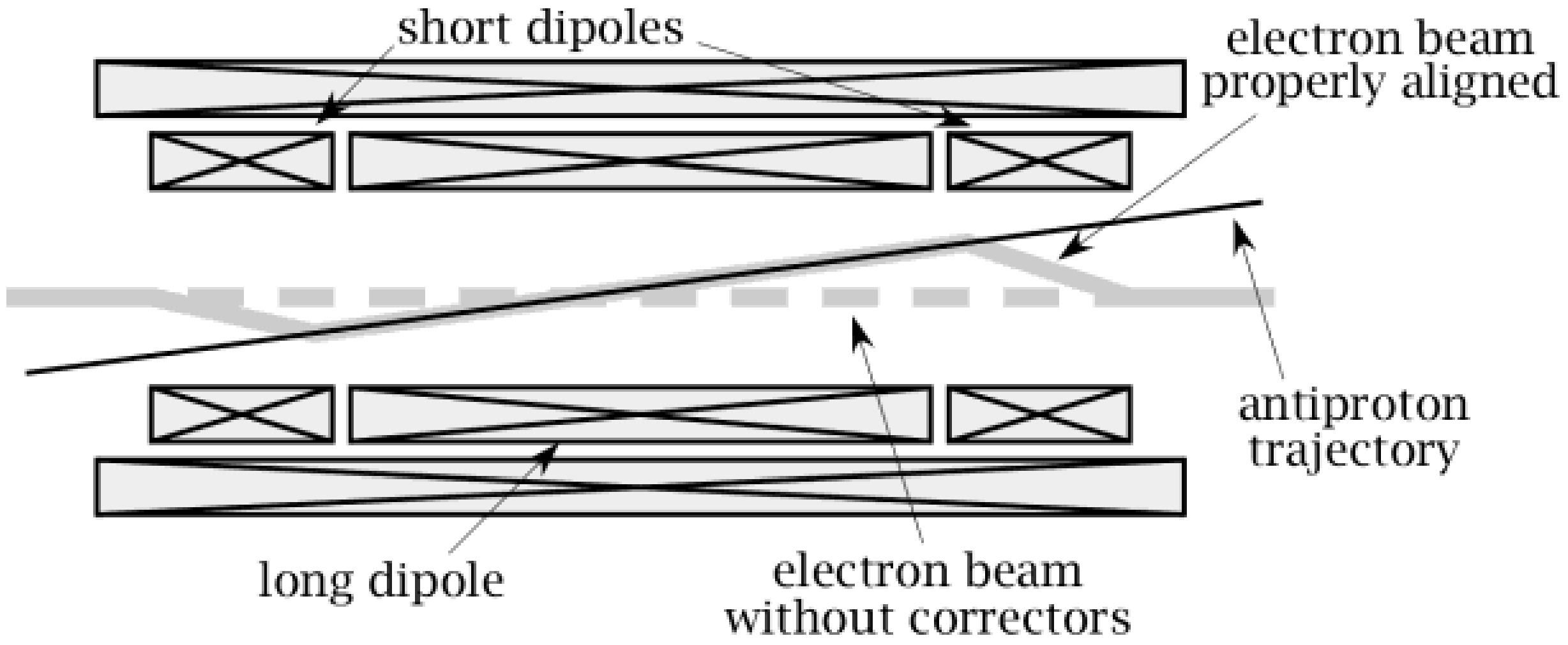

Figure 4. Sketch of the placement and action of the dipole correctors (transverse scale exaggerated). Without activating the correctors, the electron beam would follow the dashed path in the main solenoid. By using the correctors, the electron beam can overlay the (anti)proton path.

The dipoles are placed on the outer surface of the SC solenoid coil, as shown in Fig.2. Four pairs of 250-mm long coils form short vertical and horizontal dipoles at each end of the solenoid. Two pairs of 2-meter long coils are placed in the central region of the SC solenoid. The steering dipoles are wound of cable transposed from 8 wires of 0.3-mm diameter. The wire has 50 A critical current at 4.2 K and 5 T and Cu/SC ratio of 1.5. Dimensions of bare cable are $0.45\times1.48$ mm$^2$. The cable is wrapped by polyamide film of 0.03-mm thickness with 1/3 overlap. The central dipoles have one layer; lateral dipoles consist of two layers and inter-layer spacer of 0.2-mm thickness. The specific location, and the magnetic length, of each of these correctors is shown in Fig.5. The dashed line illustrates the main solenoid field on axis as a function of longitudinal position. It is at a maximum nearly from -100 cm to +100 cm and rapidly falls to almost zero at -150cm and +150cm. The solid lines in Fig.5 represent the measured strength of each set of dipole correctors. The upstream short correctors peaks around -115 cm, the long corrector extends from -75cm to +75cm, and the downstream short corrector peaks at +115 cm. Since the strength of each corrector and the main solenoid can be arbitrarily set, their magnitudes are all normalized to 1.0 in Fig.5. In the actual measurements, the solenoid was set to 6.5T, the short correctors were set to 0.8T, and the long were at 0.2T. No significant difference was observed between horizontal and vertical correctors.

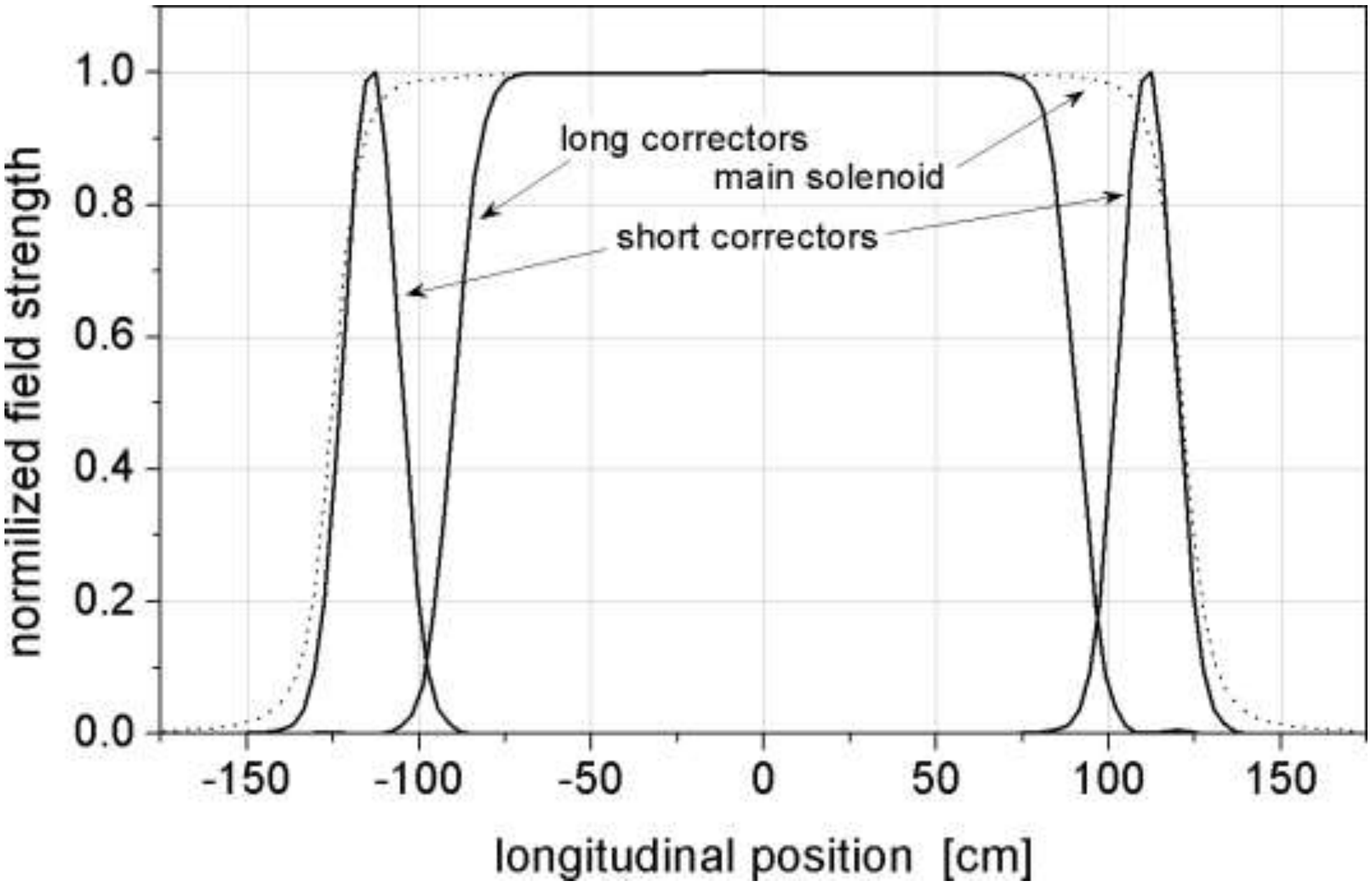

Figure 5. Normalized strength of the dipole correctors ($B_{x,y} / B_{max}$) inside the main solenoid. The solenoid field $B_z$ is also included (dashed line). The longitudinal position is referenced to the geometrical center of the main solenoid. Maximum magnetic field is 0.8T in the short correctors, 0.2T in the long correctors and 6.5T in the main solenoid.

Strongly magnetized electrons spiral around the solenoidal field lines in the TELs. The dipole correctors add a small perturbation to the nominally longitudinal solenoidal field. By superposition, the vector field of the correctors gets added to the vector field of the main solenoid.

Since the former is a uniform field pointing transversely and the latter is a uniform field pointing longitudinally, the net result is a field that points at an angle represented by the sum of the two vectors. The electron beam, following the field lines, tracks the resultant field. Beyond the region of the corrector, the field lines and the electron beam again point longitudinally, but from this new position. The total horizontal deflection *dx* can be derived from:

$$dx = \int_s \frac{B_{horizontal}(z)}{B_{main}(z)} dz \qquad (3),$$

where the two field strengths are functions of the longitudinal position *z*, and the integral covers the pertinent length shown in Fig.5. A similar expression can be written for the vertical corrector, assuming the appropriate corrector field is used. The strength of the short correctors in units of kGmm/A, which is $dx\ B_{main}$ per unit current, is about 6kG-mm/A. The strength of the long correctors is about 36 kG-mm/A. Dividing these numbers by the main solenoid strength yields a valid transverse displacement for a known amount of current. For example, short coils energized by 200A current can move the electron beam by 40 mm in 3T main solenoid field, long coils are able to deflect the beam trajectory in the main solenoid by 30 mrad if energized by 50A. Separate measurements of electron-beam deflection using BPM readings verified these calibration numbers.

### *2.3 Cryogenics and quench protection*

All the superconducting coils of the main solenoid are immersed in a liquid helium bath, and the total weight of this cold mass is 1350kg. Due to hysteresis effects and eddy currents in steel, a small amount of heat is generated whenever the current in the superconductor is changed, that limits the maximum current ramp rate to 10A/s. In practice, the main solenoid is rarely powered up or down, but usual ramp rates lie under 4A/sec. Total static heat load onto the helium vessel is 12 W and 25 W onto the nitrogen thermal shield of the cryostat. TEL cryostat is a part of the Tevatron magnet string cooling system which delivers 24 g/s of liquid helium. The magnet temperature margin equals 0.6 K at helium temperature of 4.6 K.

On a rare occasion, the TEL main solenoid has quenched, either on its own or in response to other Tevatron SC magnets quenching first. Since the number of times this has happened is extremely small, it is not considered a liability to the performance of the TEL or the Tevatron. Nevertheless, quench protection is an important subsystem of the TEL, as the main solenoid can contain up to 1MJ of energy when it reaches its maximum rating of 6.5T, and that energy is released over a mere two seconds when the solenoid is quenching. The current in each SC magnet loops through external power supplies, allowing external quench detection circuits and loads to absorb most of that energy. Simulations of quenches suggests that roughly 90% of the total energy can be dissipated in external resistive loads, with the remaining 10% being dissipated in the solenoid itself. In these simulations, the temperature of the hottest point in the coil rises to about 270K.

The dipole correctors only contain up to 1.3kJ of energy, and dissipating this energy within the magnet is not worrisome. However, heat in one region could cause a quench in the main solenoid. Therefore, the correctors are also connected to quench protection circuits and loads.

Each monitor was originally designed to observe the voltage across its magnet and the time-derivative of the current, which were compared to an assigned limiting voltage :

$$\left| L_{magnet} \frac{dI}{dt} - V(t) \right| < V_{\mathrm{lim}it} \qquad (4),$$

where $L_{magnet}$ is the inductance of the magnet. If the difference exceeds $V_{limit}$=1 V , the magnet is assumed to have begun to quench, a signal is sent to high current IGBT switches to disconnect the coil from the power supply and allow to dump the coil current into the resistive load. Mechanical current breakers are installed in series with the solid state switches for redundancy. However, the inductance of a large 0.5 H solenoid is typically not constant at low frequencies (1-10 Hz), due to iron-saturation effects and eddy currents. The overly simplified model expressed in Eq.(4) led to occasional false quench detections. A more sophisticated model using higher-order effects of both *V(t)* and *I(t)* was adopted for operation. The quench protection monitor now tests the relation,

$$\left| L_{magnet} \left(\frac{dI}{dt} + \kappa_1 \frac{d^2 I}{dt^2}\right) - \left(V(t) + \kappa_2 \frac{dV}{dt}\right) \right| < V_{\mathrm{lim}it} \qquad (5).$$

The addition of extra terms in Eq.(5) offers the ability to better mimic the physical behavior of the magnets over a range of frequencies that has dramatically decreased the number of false quench detections.

The power supplies for each of the solenoids and correctors need to be able to sustain each magnet's full current. The main solenoid in normal operation requires a full kilo-Ampere; large MCM500 cables carry this current over some 60 meters from the power supply (located in an above ground gallery) to the solenoid itself located in the Tevatron tunnel. The short dipole correctors employ 200-A supplies, while the long correctors use 50-A supplies. Since the correctors might need to be energized in either direction, each supply is fed through a reversing-switch box. This box is able to swap the leads, effectively turning the unipolar supplies into bipolar supplies.
The current ramp rates for each of the superconducting magnets are limited, and all of the settings are done remotely through computer control. The reversing-switch circuits automatically handle ramping the current through zero and switching polarity properly. With this feature, scanning the electron beam transversely becomes straightforward.

## *2.4 Straightness of field lines*

If there is any significant deviation of the magnetic field lines from a straight line of (anti)proton trajectory, then the electron beam, which follows the field lines, would not interact properly with the (antiproton) bunches. Both initial design considerations [2] and operational experience [3,4] emphasize the need of the solenoid field straightness within 0.2 mm, that is a small fraction of the (anti)proton rms beam size $\sigma$=0.5-0.7 mm and the electron beam radius $a_e$=1.5-2 mm in the TEL. The electron lens magnets were designed and built to be straight and uniform within the specification that was later confirmed in special measurements of the magnetic field lines.
The measurement technique is illustrated in Fig.6. A small iron rod was centered in a non-magnetic gimbal equipped with low friction sapphire bearings and mounted on a small cart. The cart was dragged through the solenoid, and the solenoid field magnetized the rod, which aligns itself along the field lines (a magnetized ferromagnet feels a torque $\boldsymbol{M}\times\boldsymbol{B}$ attempting to align it along the field lines). A small mirror that was attached perpendicularly to the rod (actually surrounding the rod) reflected a laser beam from one end of the solenoid back down the same direction. The reflected laser beam (returning at twice the angle of the magnetic rod) struck a two-dimensional light-position sensitive detector (PSD). This large PSD was read out by the processing electronics that directly reports the XY coordinates of the incident light spot. In this manner, minute angles $\theta_{x,y}=B_{x,y}/B_{main}$ of the order of few microradians could be observed. The field line coordinates *(x,y)* are then calculated as:

$$(x,y)=\int_z \theta_{x,y}dz \qquad (6).$$

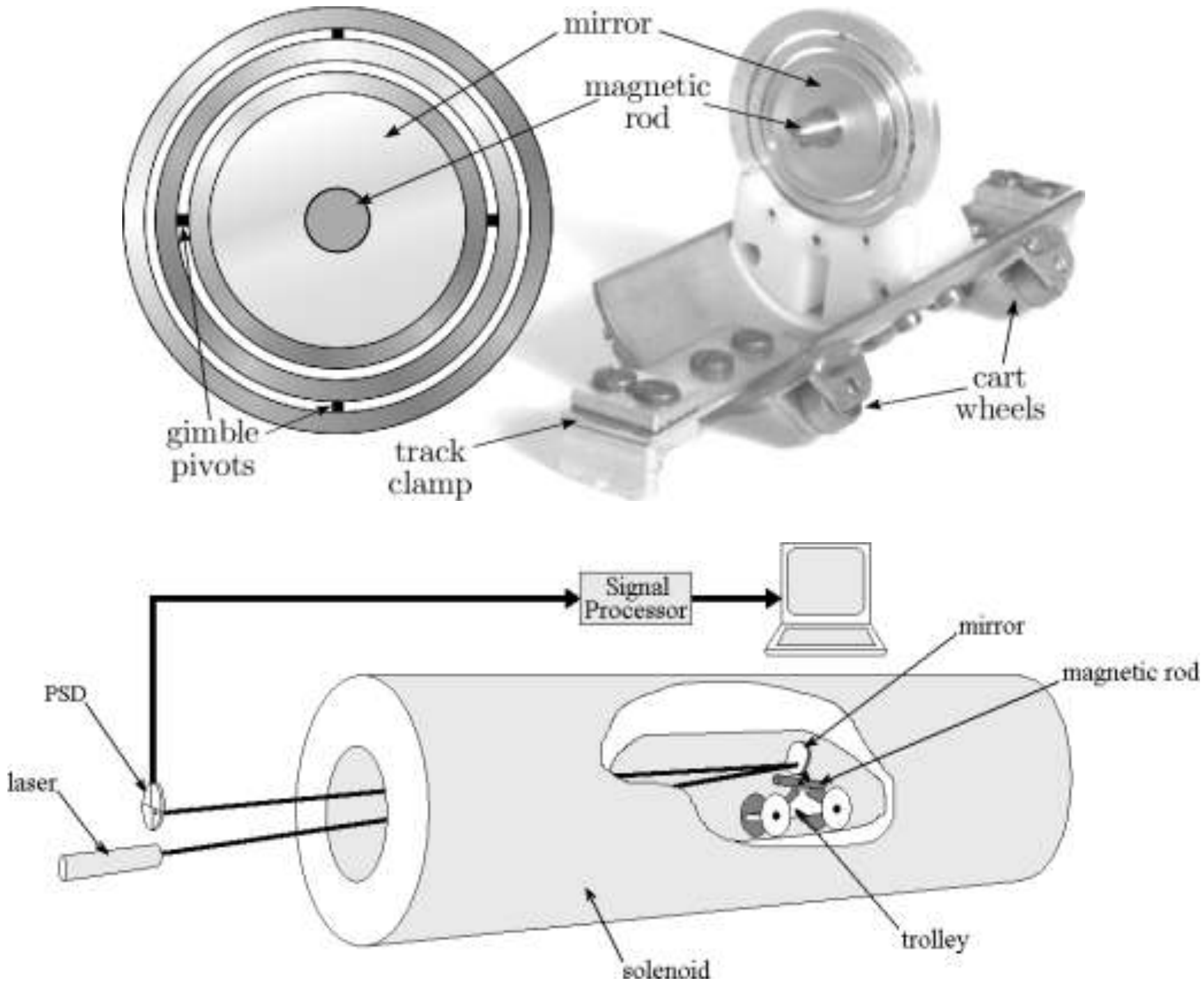

Figure 6: (top) Drawing of the gimble and photograph of the cart. The gimble measured only an inch across and has very little mechanical resistance. The cart was pulled by a long track, and it rolled inside an aluminum pie temporarily thrust into the solenoid bore; (bottom) simplified cartoon illustrating the technique used to measure the magnetic field line straightness in the main solenoid.

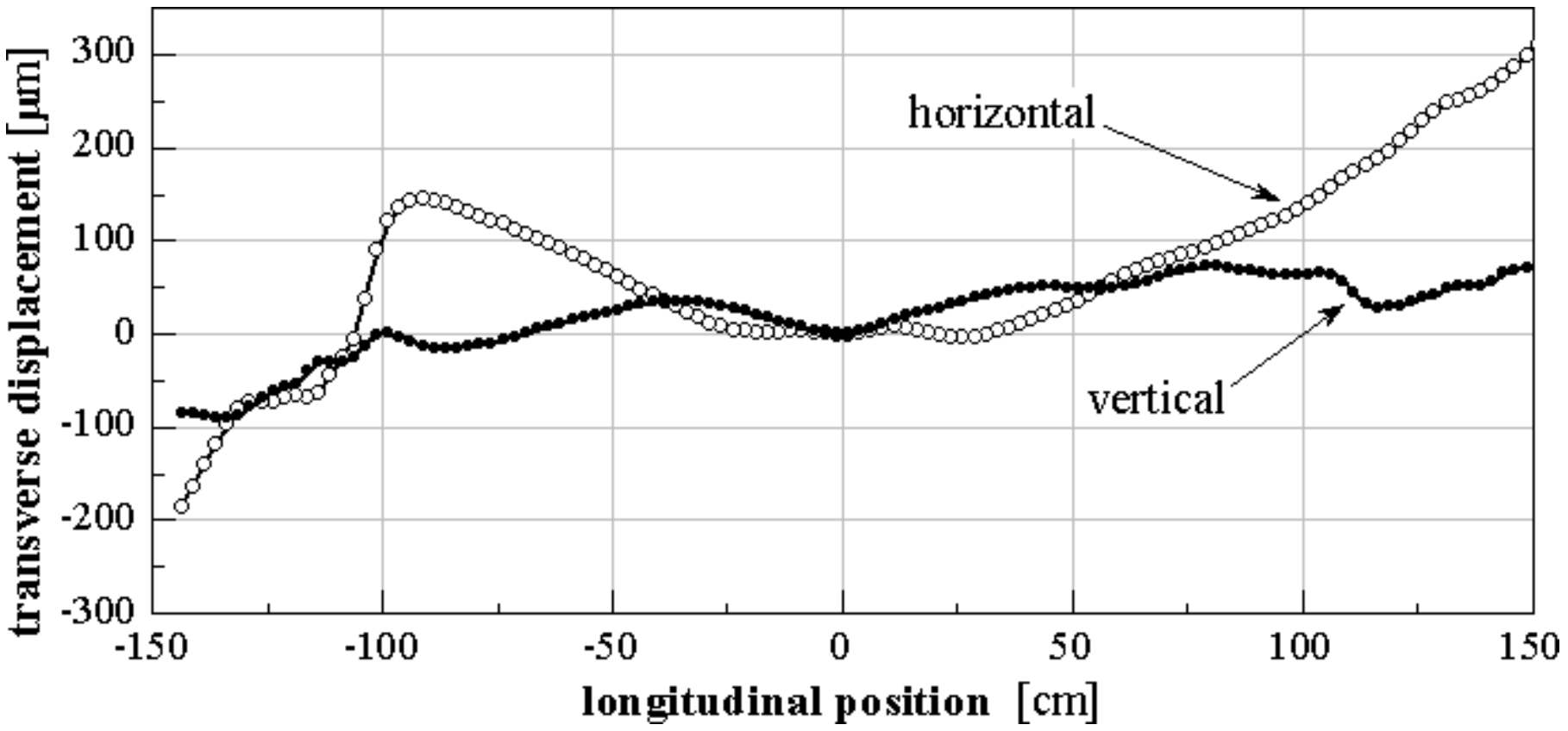

Figure 7: Measured vertical and horizontal field lines in the TEL-1 main solenoid at 4T.

The field lines in the center of the solenoid are shown in Fig.7. In the +-100 cm of the interaction region, the field does not bend more than 200 μm in the horizontal direction and only about 45 μm in the vertical. The rms deviations of the magnetic lines are 15 μm in the vertical plane and 50 μm in the horizontal plane. Therefore the electron beam is able to surround the antiproton bunches through the entire solenoid length. The 200μm variation is conveniently small and allows the option to experiment with beam-beam compensation at different electron-beam sizes.

The deviation of the straightness of the field lines change as the solenoid's field is ramped up or down is found not exceeding 20μm. The solenoid field lines as far as +- 1 mm away from each other are measured to be parallel within +-6μm [10].

# 3. Electron beam system

Following Eq.(1), electron current densities of the order of $j_e$=50 A/ cm$^2$ in a 10kV electron beam are required in order to shift the tune of 980 GeV (anti)protons by $dQ$=0.01 (assuming a 2 meter long electron lens at the location with $\beta_{x,y}=100$m). Below we describe electron guns and collectors that have been developed to generate and collect the TEL electron beam.

## *3.1 Electron guns*

Magnetic system of the TEL allows adiabatic magnetic compression of the electron beam cross section area by a factor of $(a_c/a_e)^2=B_{main}/B_{gun}$ ~10, so the required maximum electron current density at the cathode is about $j_c= j_e/(B_{main}/B_{gun}) \approx 5$ A/ cm$^2$. In order to have the electron beam radius $a_e$ to be several times the rms (anti)proton beam size $\sigma$ in the TEL, the cathode radius should be $a_c$=5-10 mm Because of significant differences in dynamics of different Tevatron bunches [1], the electron beam of the electron lens has to be modulated with high duty factor and characteristic on-off time of about 0.5-1 microsecond. High current density, fast modulation and requirement of smooth current density profile led to choice of the electron gun with convex cathode which allows to get higher perveance and perform current modulation by anode voltage (i.e., no grid). During experimental beam studies in the Tevatron – described in detail in [4] - several electron current profiles were found to be effective: a) rectangular distribution results in uniform tune shift $dQ$ for all high energy particles passing through the electron beam, but has disadvantage of strong nonlinear space charge forces beyond boundaries of the electron beam; b) bell-shape (close to a Gaussian) distribution has weaker nonlinearities but smaller beam size as well, that makes it somewhat cumbersome to align it on the beam of (anti)protons; c) “smooth edge and flat top” (SEFT) distribution that combines advantages of both previously mentioned distributions.

Correspondingly, three electron guns have been developed for the TELs. One of the most important characteristics of an electron gun is its perveance $P$:

$$P=I/U_a^{3/2} \qquad (7),$$

where $I$ is the beam current and $U_a$ is the anode potential with respect to the cathode. For the guns with flat or concave cathodes, current density inhomogeneity becomes large when the perveance exceeds the value of 1 – 2 $\mu$A/V$^{3/2}$. In the case, where the gun has to be immersed into a strong longitudinal magnetic field, the perveance can be increased by usage of a convex cathode [11].

The electron guns were simulated and optimised using UltraSAM code [13] in order to have the desired current density distribution and high perveance. The geometries of the guns are shown in Fig.8 together with electric field distribution along the beam axis (the guns have axial symmetry) and electron trajectories. The guns employ spherical cathodes with +-45 deg opening angle. A Pierce type electrode (“control electrode”) is installed around the cathode of the ”flat” gun (Fig.8a) for manipulation of the beam current density distribution. Control electrodes of different geometry are installed for the same purpose between cathode and anode in the “Gaussian” and around the cathode in the SEFT guns (Figs. 8b and 8c). The control electrodes in the latter two guns are usually kept at the same potential as the cathode.

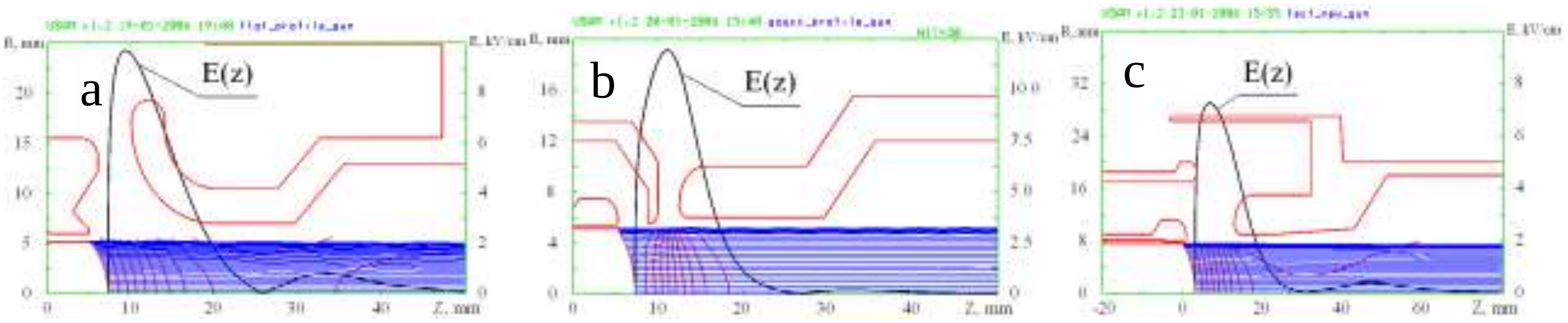

Figure 8: Gun geometry and “UltraSAM” code electric field simulation results for: a) “flat-top” gun; b) Gaussian gun; c) SEFT gun.

Mechanically, all three guns look similar as shown in Fig. 9a. They are assembled on a 6 ¾ inches diameter stainless steel vacuum flange and use ceramic rings as insulators between electrodes. The guns employ spherical convex dispenser cathodes purchased from HeatWave Labs (Watsonville, CA). 10 or 15 mm diameter barium-impregnated tungsten cathodes operate at temperatures 950-1200°C. They are equipped with Mo-Re support sleeve and molybdenum mounting flange and have internal heater filament (bifilar option, one heater lead internally grounded). Near-cathode electrodes are made of molybdenum, while the control electrodes and anodes are made of oxygen free copper.

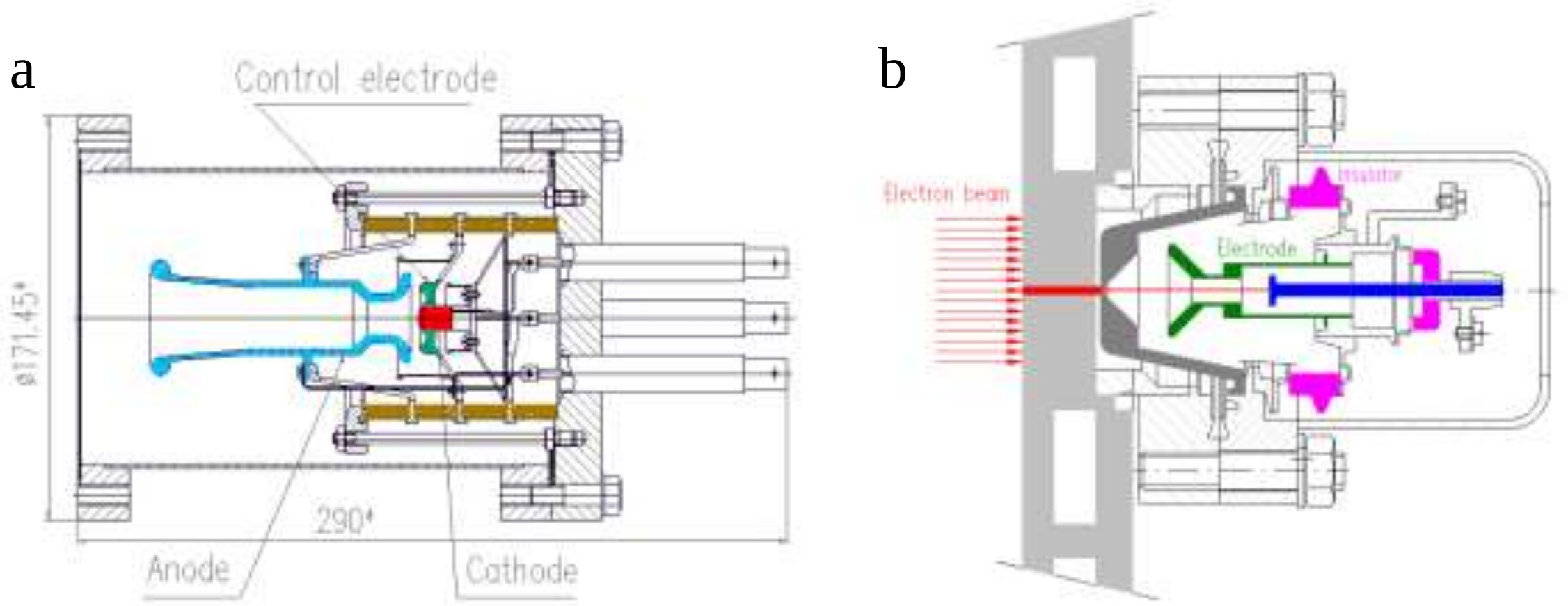

Figure 9: a) Mechanical design of the “flat-top” gun; b) pin-hole collector assembly for beam profile measurements on the test bench.

Gun characteristics were measured on the test bench used at Fermilab for prototyping the TEL elements [12]. The test bench consists of the gun immersed into longitudinal magnetic field $B_{gun}$ of 1- 2 kG generated by a gun solenoid, a drift tube with diagnostics placed inside 4 kG, 2 m long main solenoid, and a collector, also inside a separate solenoid. The collector is equipped with a beam analyzer, illustrated in Fig. 9b. A small 0.2 mm diameter hole in the collector base lets a narrow part of the electron beam to pass through a retarding electrode and be absorbed by an analyzer collector. To measure the transverse current density distribution, the beam is moved across the hole by the steering coils installed inside the main solenoid, and the analyzer collector current is recorded as a function of the transverse beam position.

Except for their high perveance, the guns are not much different from a planar cathode gun. The beam currents follow the Child's law with a good precision (Fig.10) yielding perveances of 5.3 $\mu A/V^{3/2}$, 4.3 $\mu A/V^{3/2}$, 1.8 $\mu A/V^{3/2}$ for the "flat-top", SEFT and "Gaussian" guns, respectively. To prevent thermal problems at the collector, the total current and profile measurements were done in the DC regime at currents below 0.5-1 A. The gun characteristics at higher currents were investigated in a pulsed regime with the pulse width of 0.2-4 μs. No significant deviation from the results of the DC measurements was found.

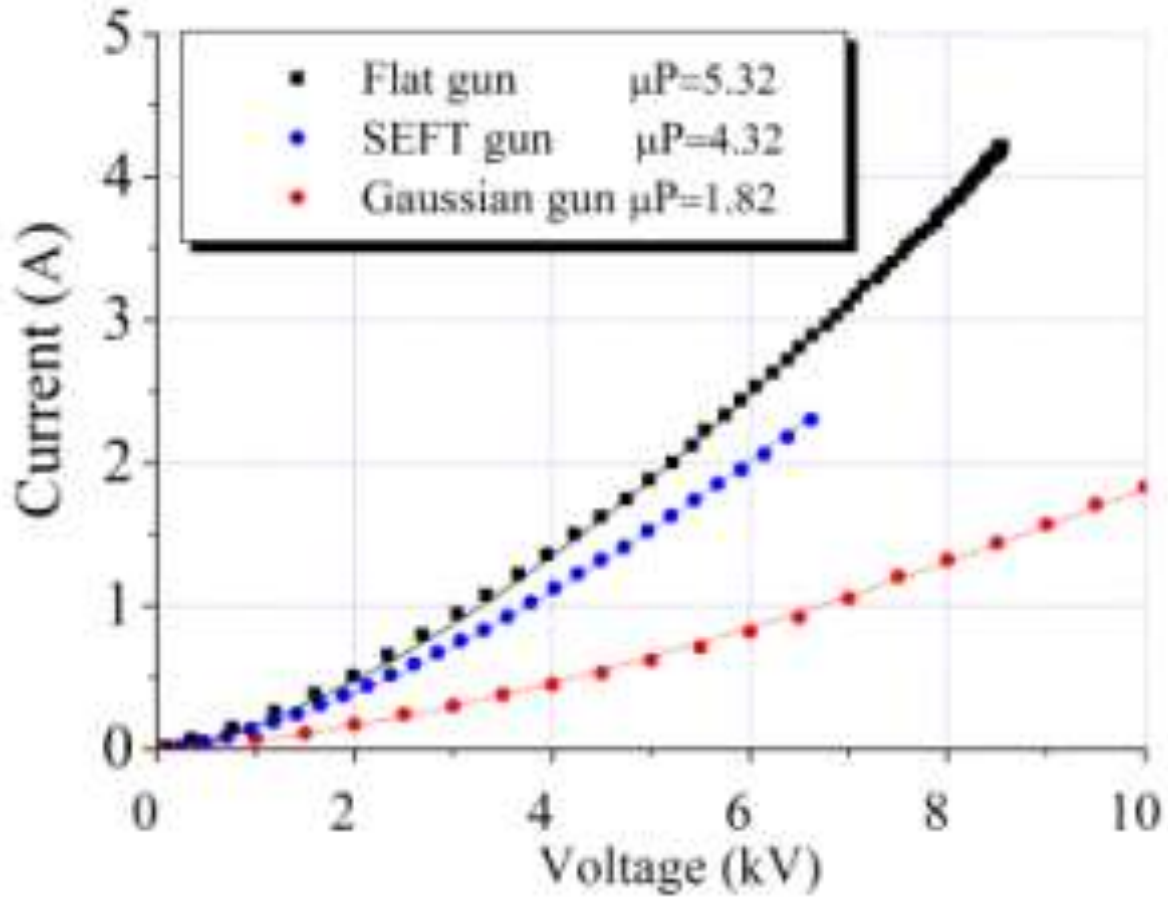

Figure 10: Volt-Ampere characteristics of the three electron guns, solid lines are fits according to Child's law $P=I/U_a^{3/2}$.

An example of the 2D profile of the electron current distribution from the SEFT gun measured by the "pin-hole" collector is shown in Fig.11a. Current density variations are less than 10% over 90% of beam diameter. Measured and calculated one dimensional profiles of electron beams from all three guns are presented in Fig.11b. There is a good agreement between predicted and observed current densities over the most of the beam area except the very edge of the beam.

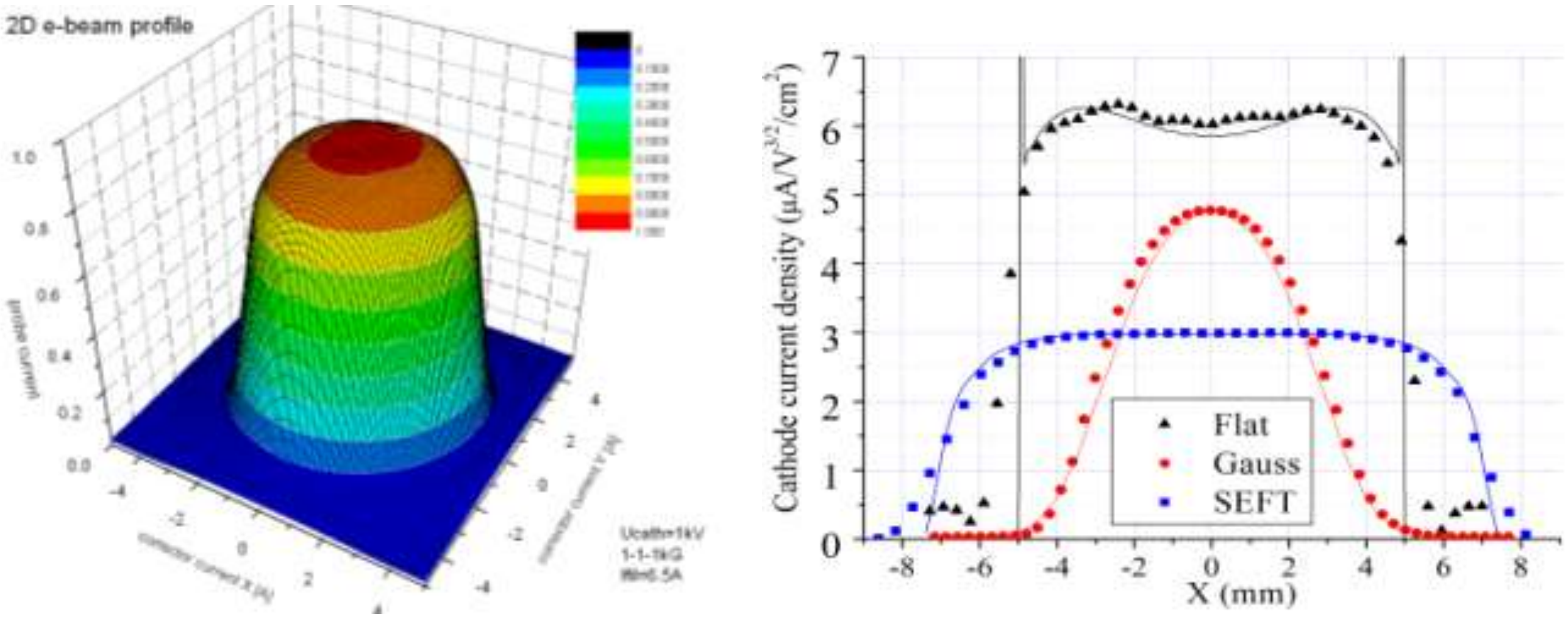

Figure 11: a) (left) 2D electron current density distribution of the SEFT gun beam; b) (right) 1D current density distributions for three guns, solid lines represent UltraSAM simulation results. In both cases, the control electrode voltage was set equal to the cathode voltage.

Electron emission from the edges of the cathode is strongly dependent on the accuracy of the mechanical alignment of the near cathode or the control electrodes w.r.t. the cathode. Fig. 12 shows 1D current profile in the case when the control electrode of the flat gun was (unintentionally) set a bit farther from the anode than the cathode. The edge peaks in the current density profile indicate, and computer simulations confirm, that the reason is some 0.4 mm protrusion of the emitting surface from the near-cathode electrode with respect to its optimum position. The shift occurred because of either uncertainty in the thermal expansion of the cathode or mechanical error. Probably, a slight current distribution asymmetry, seen in Fig.12, is because of an asymmetric misalignment. Application of negative voltages to the near-cathode electrode (with respect to the cathode potential) resulted in suppression of the electron emission at the edge and can lead to narrower (almost bell-shape) current profile, as illustrated by Fig.12. Total beam current reduction factors for the SEFT and the “flat-top” gun are shown on Fig.13 as functions of (negative) profile control voltage $U_{pr}$ normalized to (positive) anode-cathode voltage difference $U_{ac}$. Application of positive voltage to the profile control electrode results in formation of hollow beam which is disadvantageous for the beam-beam compensation purposes but can be used effectively for other applications, such as electron cooling [14] or collimation [15]. In conclusion, the current density distribution (profiles) in the TELs can be changed by employing different geometries of the gun electrodes or can be varied by voltage on the profile controlling electrodes.

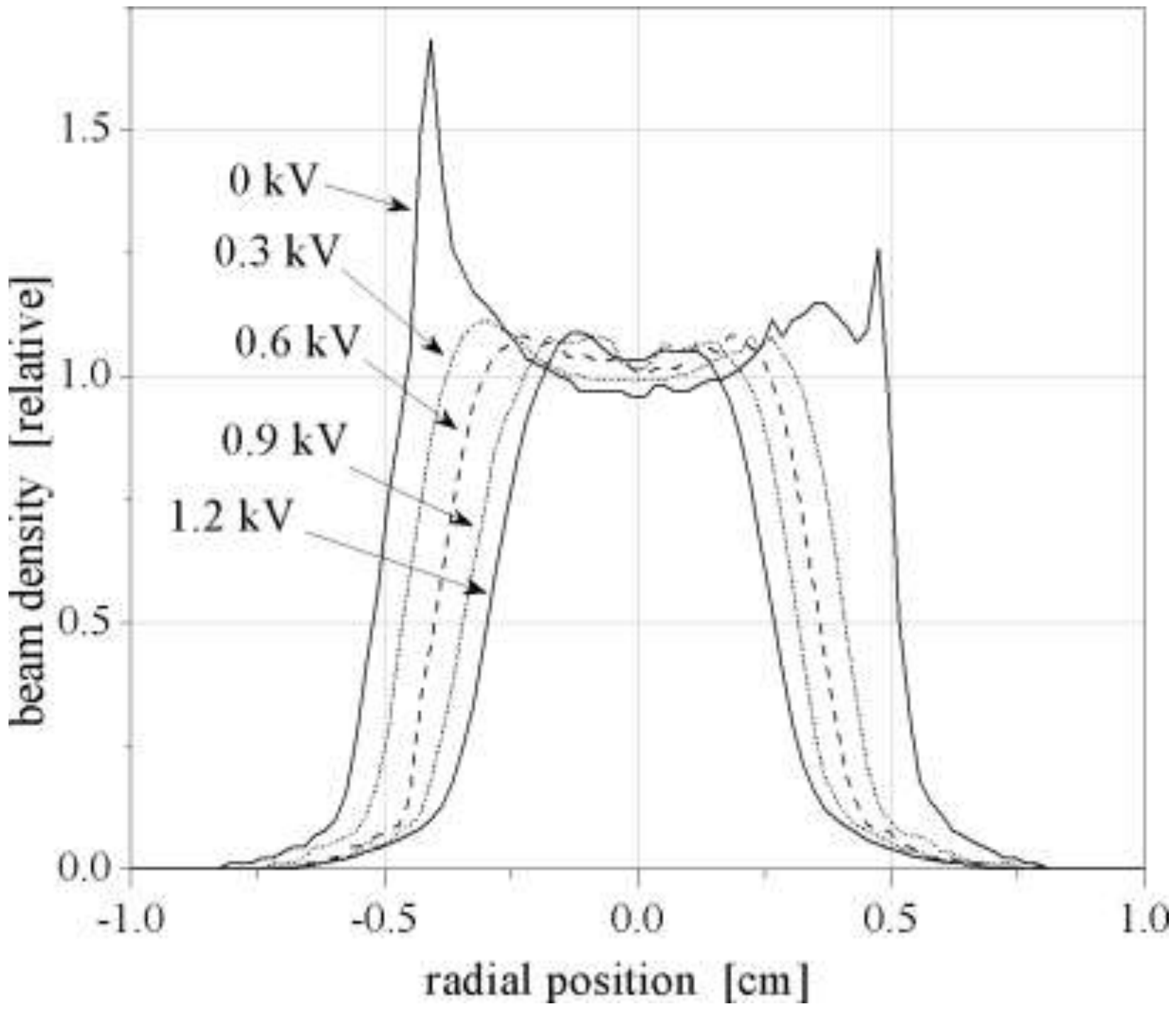

Figure 12: The electron current density profile generated by the "flat" gun at different voltages $U_{pr}$ on the near-cathode electrode. Anode-cathode voltage $U_{ac}$= 3 kV, magnetic field in all solenoids 2 kG.

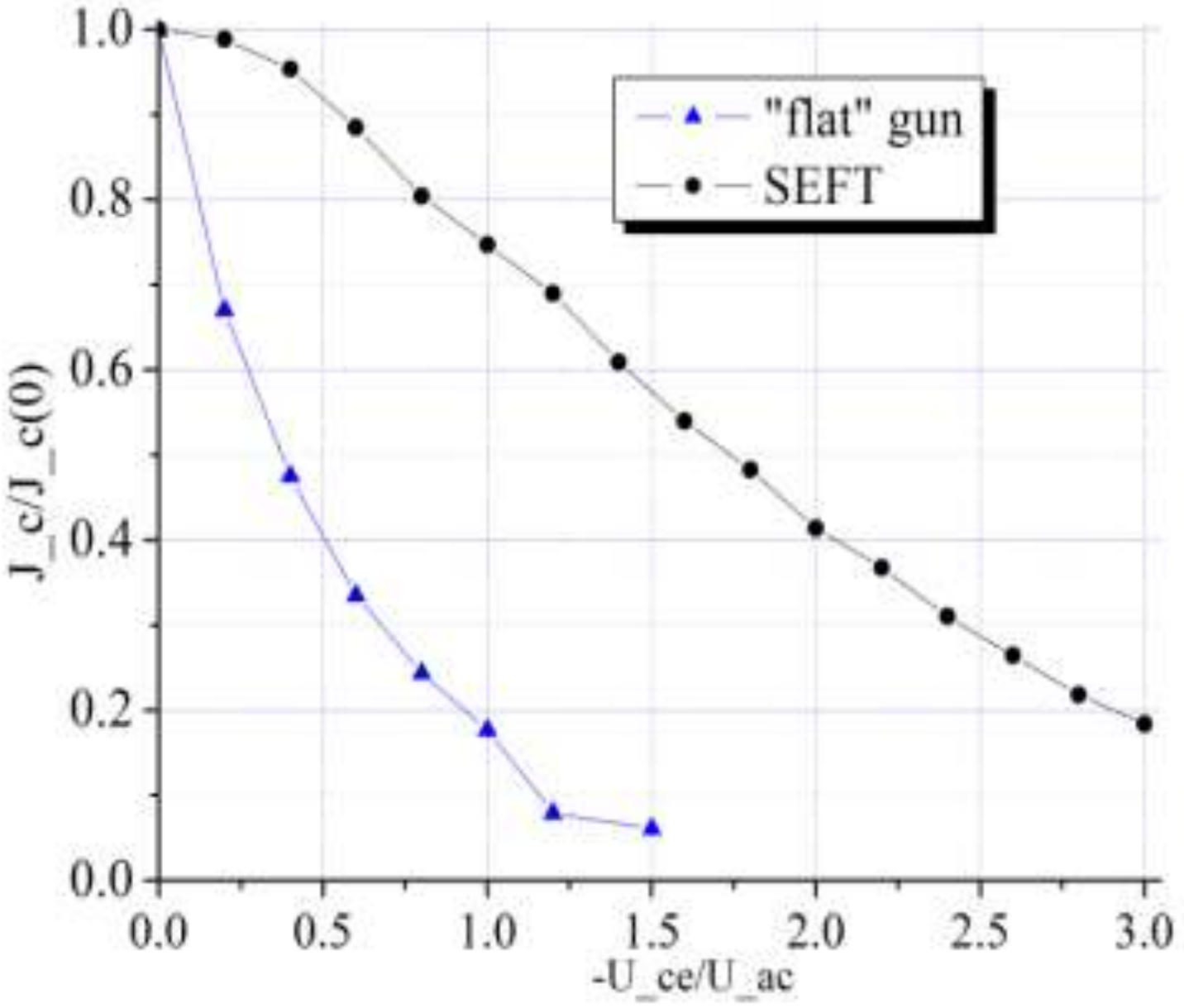

Figure 13: Total current of the "flat" and SEFT guns vs profile controlling voltage.

Total filament power required to keep the cathode at operational temperature of 1000-1100 deg C is about 35-45 W for 10 mm diameter cathode (used in the “flat” and “Gaussian” guns) and about 60-70 W for the 15 mm diameter cathode used in the SEFT gun (see Table II). For initial activation of the cathode, the power is increased by 30-50% for a short period of time until the cathode starts to generate enough current to follow the Child’s law (Eq. 7) at the design cathode-anode voltages. Special care is taken in order to have sufficiently good vacuum in the gun area in order not to poison the cathode that may result in reduced cathode lifetime. With all the precautions, the cathodes of the guns installed in the TELs operate for several years without significant deterioration. If the gun is exposed to air, at high cathode temperatures the cathode is ruined (the tungsten gets oxidized, creating a layer with a high work function) and either a complicated cathode surface processing or (easier) the cathode replacement is needed.

## *3.2 Electron beam collector*

A sketch of the electron beam collector is shown in Fig.14, with a view of the expanding electron beam. Spreading out the beam has several advantages, one being the distribution of the heat load. Locating the heating in one spot could melt the copper if the TEL is operated at full beam power (3-5A at 10kV). Instead, the magnetic field lines beyond the collector solenoid spread out, and the electrons, following the field lines, are absorbed by a much larger area of the copper collector. Additionally, chilled water is piped into the collector, where it passes through ducts within the copper and extracts up to 50 kW of heat. The intake pipe is shown in Fig.14 to illustrate the setup. Another serious concern is the production of secondary electrons whenever an energetic primary electrons impinges on the surface. The secondary electrons penetrating back into the main solenoid can adversely interact with the primary electron beam, generating a two-stream instability even if magnetized [16].

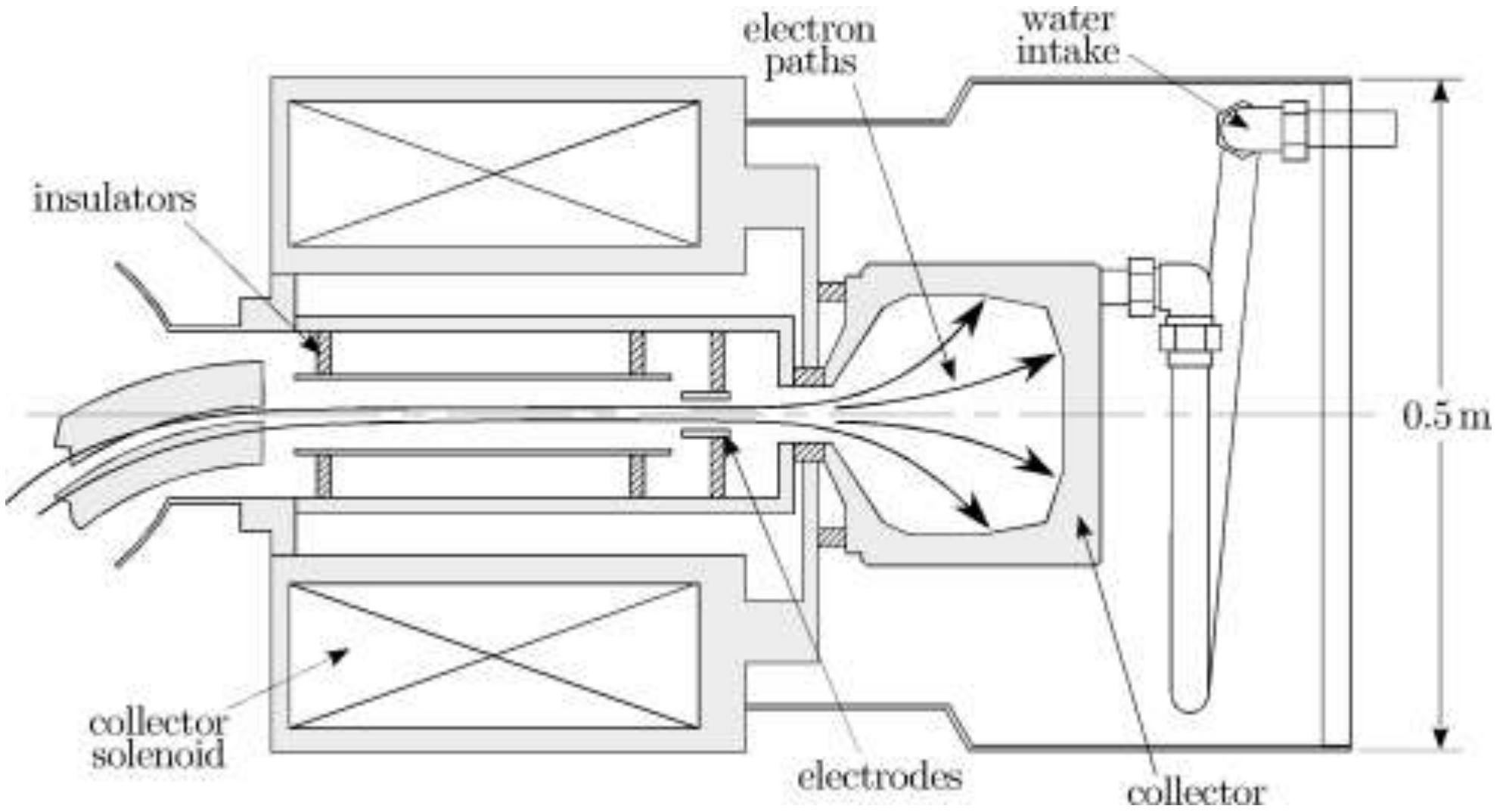

Figure 14: Scaled sketch of the collector cross-section. The collector itself is a water-cooled copper cavity that resides outside the solenoid. This allows the electron beam to spread out, distributing the heat load and decreasing the production of secondary electrons.

The design of the collector targets this issue by employing "magnetic bottle" principle – only electrons with small enough transverse velocity can travel from a region with low-magnetic field *B(0)* to higher magnetic field *B(z):*

$$v_{\perp}^{2} < \frac{v_{||}^{2}}{\left(\frac{B(s)}{B(0)} - 1\right)} \qquad (8).$$

An electron must originate with enough parallel momentum to overcome the magnetic compression; if the electron does not have enough, it will run out of longitudinal momentum and be returned back to the collector surface. The TEL's collector surface has a residual field on the order of *B(0)*~100 G, while the collector solenoid runs at 3.8 kG, implying that the longitudinal momentum needs to be over six times larger than the perpendicular momentum. If the electrons are emitted from the collector surface uniformly over all solid angles, only 1.2% of those electrons meet that constraint. More importantly, the number of electrons that can pass all the way into the 35kG main solenoid is 0.14%. Experimentally, by comparison of the cathode and the cathode currents (explained in the next section), we determined that under normal operation conditions the collector is able to retain at least 99.7 % of the incident electron beam.

The collector is electrically isolated from the rest of the system. The electron beam current absorbed by the collector is brought back, via a floating power supply, to the cathode (this is detailed in the next section). The power supply voltage is positive and adjustable. Therefore, the electrons are born at the negative cathode potential and accelerated by the anode and the beam pipe, but as they approach the collector, they are slowed down to potential somewhat more positive than the cathode one. One of the advantages of this recuperation scheme is that the heat load generated by the incident beam is directly related to its kinetic energy with respect to the collector: e.g. a 10 keV electron on a -6kV collector imparts only 4keV of energy. When the TEL is operated at maximum beam power, the total power deposited in the collector could be considerable, and lowering the voltage difference is very useful.

A test of the collector consists of setting its voltage to nearly that of the cathode and measuring the current that it receives. The data from this experiment is shown in Fig.15. At zero voltage difference, only a quarter of the beam can still manage to reach the collector surface.

However, as the voltage is raised, the amount of current reaching the collector increases. As the difference approaches 1 kV, all of the current is received.

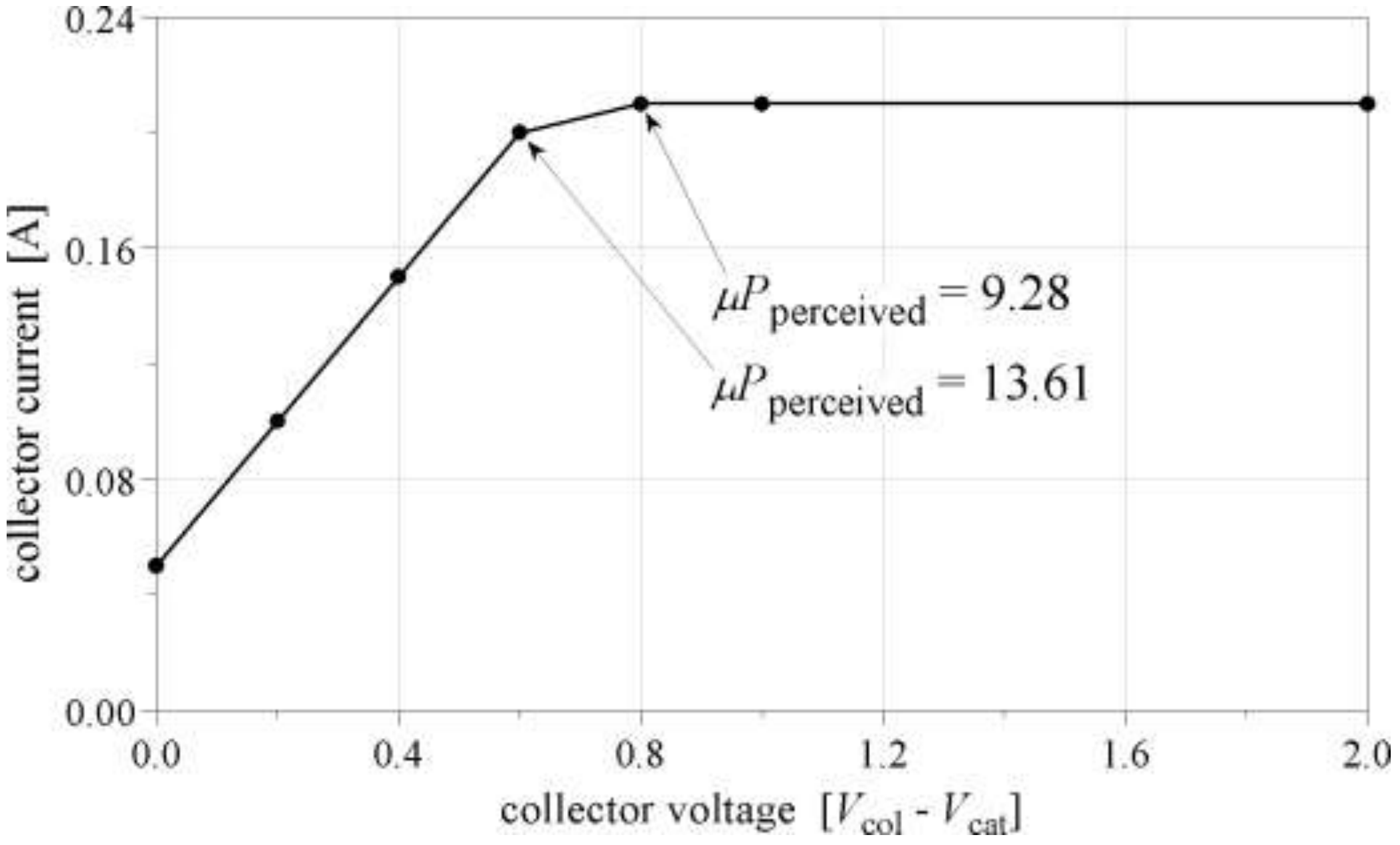

Figure 15: Measurement of the collector acceptance. As the collector voltage is adjusted with respect to the cathode voltage, the current admitted to the collector changes.

Using the data point where the maximum current is truly witnessed, the acceptance of the collector appears to be between 9.2-13.6 μP. The collector has never been the limited the TEL's performance, since its voltage can always be increased if necessary. Most commonly, the collector voltage is set at about 2-5kV above the cathode voltage. Two electrodes, which have primarily been used for monitoring the beam's passage into the collector are also shown in Fig.14. For example, the "scraper" electrode, closest to the collector, has its own current monitor, and if it reads something other than zero, the TEL beam is, at least partially, running into it. Adjusting the downstream magnetic correctors such that this signal is eliminated assures optimal performance of the TEL.

### *3.3 Electric circuit*

A recirculating electron beam could be generated by the simple circuit illustrated in Fig.16. In this case, the cathode power supply does not need to produce any current. The beam current does flow through the collector power supply, but the voltage that this supply must support can be significantly less. The anode modulator generates short positive voltage pulses but carries no DC current and simply ``kicks" the current around the loop (see discussion in the next chapter).

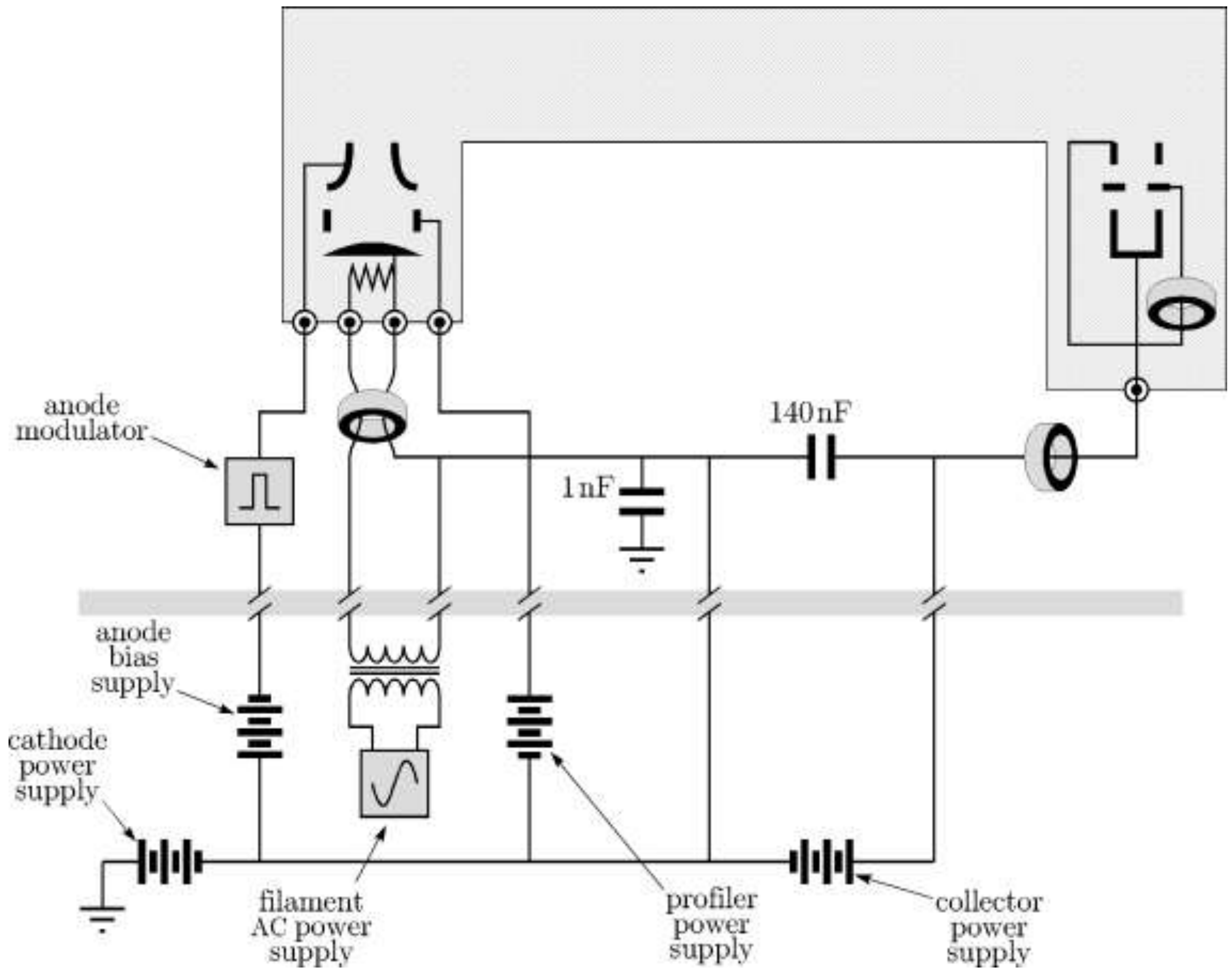

Figure 16: Schematic of the TEL electrical circuitry. The shaded barrier represents about 60 meters between the Tevatron tunnel and the gallery where supplies and electronics are stationed. The filament is at cathode potential, so the AC power supply providing the cathode heating needs to be isolated from ground via a transformer, and its signal needs to be subtracted from the cathode current meter.

The Tevatron tunnel, filled with beam induced radiation during operation, is of very limited access and a poor place for solid-state electronic equipment. Therefore, the power supplies must be located above the tunnel, connected by cables over about 60 meters in length. For the majority of these cables, high-voltage, shielded coax (RG-213) of 50Ω impedance is used. The filament is powered by 60 Hz AC which is electrically isolated through a high-voltage transformer. This allows the primary side to be referenced to ground and only the secondary is at the cathode voltage. The electron current from the cathode is measured by a high-bandwidth, commercial current transformer encircles the wire attached to the cathode. The transformer's signal is preserved over the long Heliax cable by only grounding it upstairs where an oscilloscope measures it. This prevents ground noise to corrupt the signal. Keeping the impedance at 50Ω also reduces electrical coupling from other sources over the long propagation distance and reflection issues. Rise-times of 1-2 ns and currents of a few milliAmperes are visible. However, the low-frequency filament current passes through the current transformer, so the return cable was obliged to pass through it also; the two currents are always in opposition and therefore cancel. In order to house and connect the capacitors, current meters, and make additional interconnections, a high-voltage enclosure was constructed and installed next to the electron gun.

In order to confirm that the electron beam arrives at the collector without losses, another current transformer monitors the current returning from the collector to the recirculating capacitor. A third transformer watches the scraper electrode's current since this provides the narrowest aperture and the easiest way to adjust field strengths in order to steer the beam into the collector. The scraper feeds into the collector cable, so that the collector current should be identical (though delayed) to the cathode current. Above several hundreds of mA, the peak collector current is somewhat less than that of the cathode current due to electron pulse lengthening induced by electron's own space-charge in the beam pipe.

## *3.4 Electron beam modulators*

Beam-beam effects are unique for each of the Tevatron proton or antiproton bunches [1], correspondingly, compensation of them requires different electron currents prepared for each bunch. There are 1113 RF buckets along the Tevatron orbit (RF frequency 53 MHz, revolution frequency 47.7 kHz), but only 36 of them are populated with proton bunches and 36 with antiproton bunches. For each kind of particles, the populated buckets are arranged in three trains with twelve bunches in each of them. Total length of each bunch train is ~4.5 μs. The distance between the batches is 2.6 μs. The bunch spacing within a train is 396 ns. The interaction length of the TEL is about 2 meters and it takes 33 ns for 10kV electrons ($\beta_e$ = 0.2) to traverse its length. The (anti)proton bunch enters the drift space of the lens when it is already filled with electrons. For (anti)protons ($\beta$ = 1) it takes 6 ns to pass through the TEL. After the last particle of the (anti)proton bunch leaves the drift space, it is possible to shut off the electron gun. Taking into the account also the (anti)proton bunch length of about 10 ns, the minimum flat top length requirement to the TEL gun's extraction voltage pulse is about 50 ns. The Tevatron bunches share the same beam pipe and, ideally, in order for the TEL to act on only one type of particles leaving another unaffected, the total electron pulse duration should be limited to less than 400 ns. The proton and antiproton bunch structure in the Tevatron and required timing of the electron beam in the TEL are shown in Fig.17. Besides the pulses of electron beam to compensate for beam-beam effects, additional electron pulses are required in the abort gaps between the bunch trains (not shown in the picture) for the TEL to remove uncaptured DC beam particles [5]. Therefore, the system ideally should produce 39 pulses during each period of particle revolution in the Tevatron (every 21 μs). The required electron current modulation can easily be done by a gridded gun, but such a method is not acceptable for the TELs as it destroys the carefully prepared electron current density distribution generated on the cathode. Therefore, the modulation should be done by full anode voltage of 5-10kV and that results in extremely challenging requirements for the TEL gun HV modulator. The requirements have been significantly eased by following three considerations: a) the proton and antiproton beams are spatially separated by some 8-10 mm in the area where TEL is installed and it was found that electron beam centered on one of beams produces very little harm on the other one [4] – therefore, the total required electron pulse length can be as long as 800 ns (to fit between three neighbor bunches of the same kind); b) only a few (3-6) bunches of protons suffer greatly from the beam-beam effects and need immediate compensation by the lenses [3] – and consequently, only few electron pulses per turn are needed; c) finally, it was found that functions of the two TELs can be separated – one of them could be used for the DC beam cleaning in the abort gaps while another can do beam-beam compensation – therefore, the need to generate simultaneously many different purpose electron pulses dropped [5].

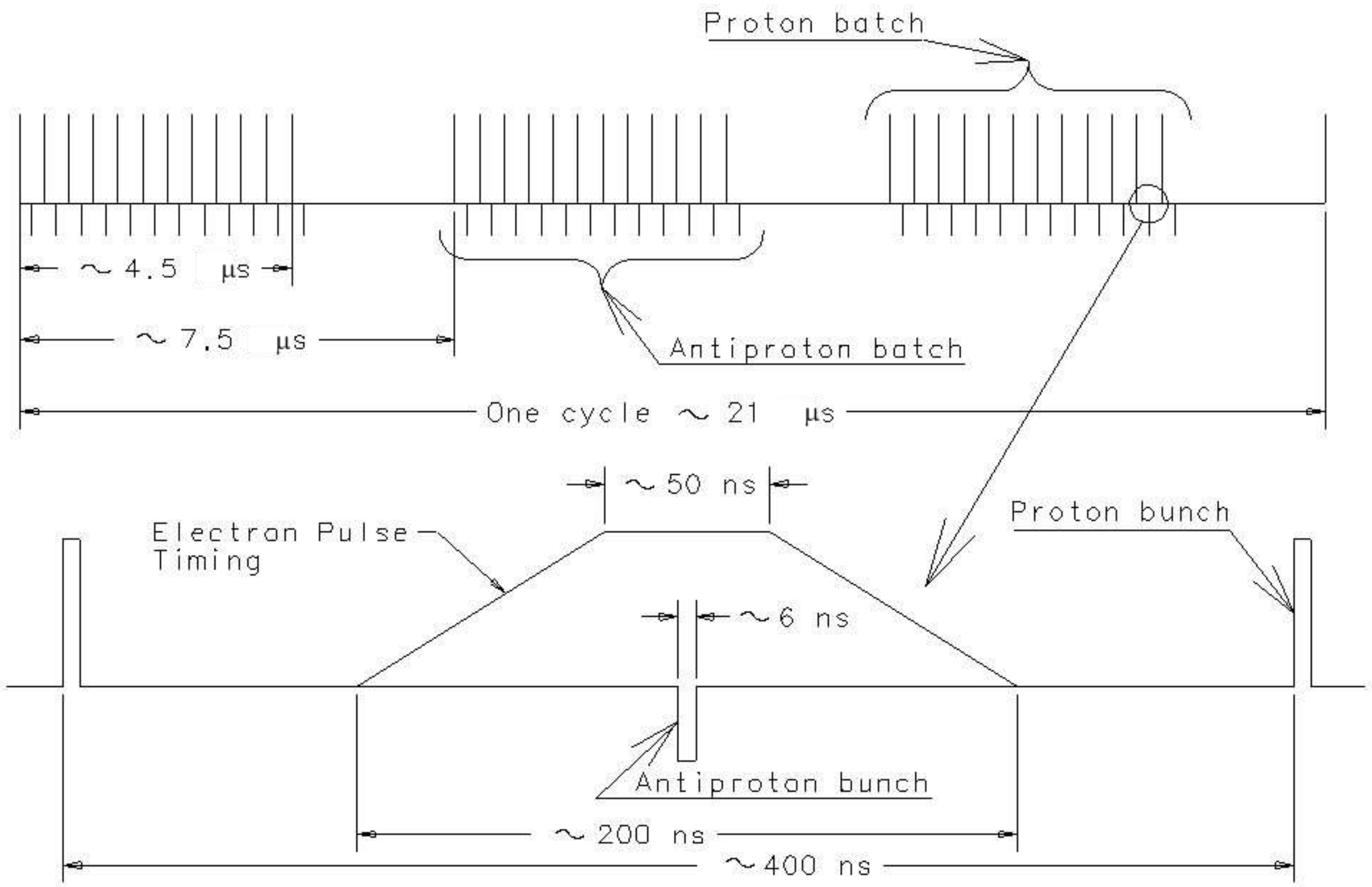

Figure 17: The Tevatron beam structure and required TEL electron pulse structure.

We have developed and tested several types of high-voltage anode modulators for the TELs [17-20]. The RF-tube based amplifier [17] and the Marx generator [19] are found most suitable for the electron lens operation in the Tevatron.

The first modulator type uses the output from the anode of a grid driven tetrode. The tube anode is connected to a +10kV DC anode supply through a 1500Ω resistor. The modulating voltage on the anode of the tetrode is then AC-coupled through two 1000pF ceramic capacitors to the electron gun anode. This modulator has the advantage that it is not susceptible to radiation damage and can be installed directly adjacent to the Tevatron beamline. A CPI/EIMAC 4cw25000B water-cooled tetrode, with a maximum plate dissipation of 25kW, is used in this modulator. Its anode voltage is supplied by a Hipotronics 10kV,16A, dc power supply. An additional LC filter (1.5H, 20μF) was added to the output of the Hipotronics supply to reduce ripple to less that 1 part in 10,000. The anode supply is connected to the tetrode through a 1500Ω, 250kW, water cooled resistor (Altronic Research). The grid of the tetrode is driven by the IGBT pulser. For compensating a single bunch of protons or antiprotons, the tube is typically operated with a screen voltage of 500V and a DC grid voltage of 0V. The tetrode's grid is then pulsed with a negative voltage pulse from the IGBT pulser, reducing the current flow through the tetrode. The positive pulse appearing on the anode is then coupled, using two 1000pF ceramic capacitors in parallel, through a short (0.6m) section of 50Ω, RG213 cable to the anode of the electron gun. Since the gun anode must be charged through the 1500Ω resistor, the risetime is limited by the sum of the tetrode's anode-screen capacitance (35pF), the capacitance of the cable connecting the modulator to the gun (60 pF), and the gun anode to ground capacitance (60pF). A typical rise- and fall-times are ~300 ns and total output pulse duration from such a modulator is 800-1200 ns [17]. A pulse to pulse amplitude stability of 0.02% was achieved by applying a feed-forward compensation signal to the grid of the tetrode to reduce ripple on the modulator output at power line frequencies. The RF-tube based modulator is in routine use in the TEL-1 since 2001.

The solid-state Marx generator drives the anode of the electron gun to produce the electron beam pulses in the second TEL. It drives the 60 pF terminal with 600 ns pulses of up to 6 kV with repetition rate of 47.7 kHz and with rise and fall times of about 150 ns. The generator consists of capacitor banks charged through inductors and erected using triggered switches, typically solid-state IGBTs. Stangenes Industries constructed ours, and each of its 12 stages can be charged to 1.2 kV and erected in series to produce a 14 kV pulse [21]. The discharge IGBT switches erect the pulse with a rise time of 150 ns, and the charging IGBT switches terminate the pulse with a fall time of 150 ns, and then provide a path for charging the capacitors in parallel.

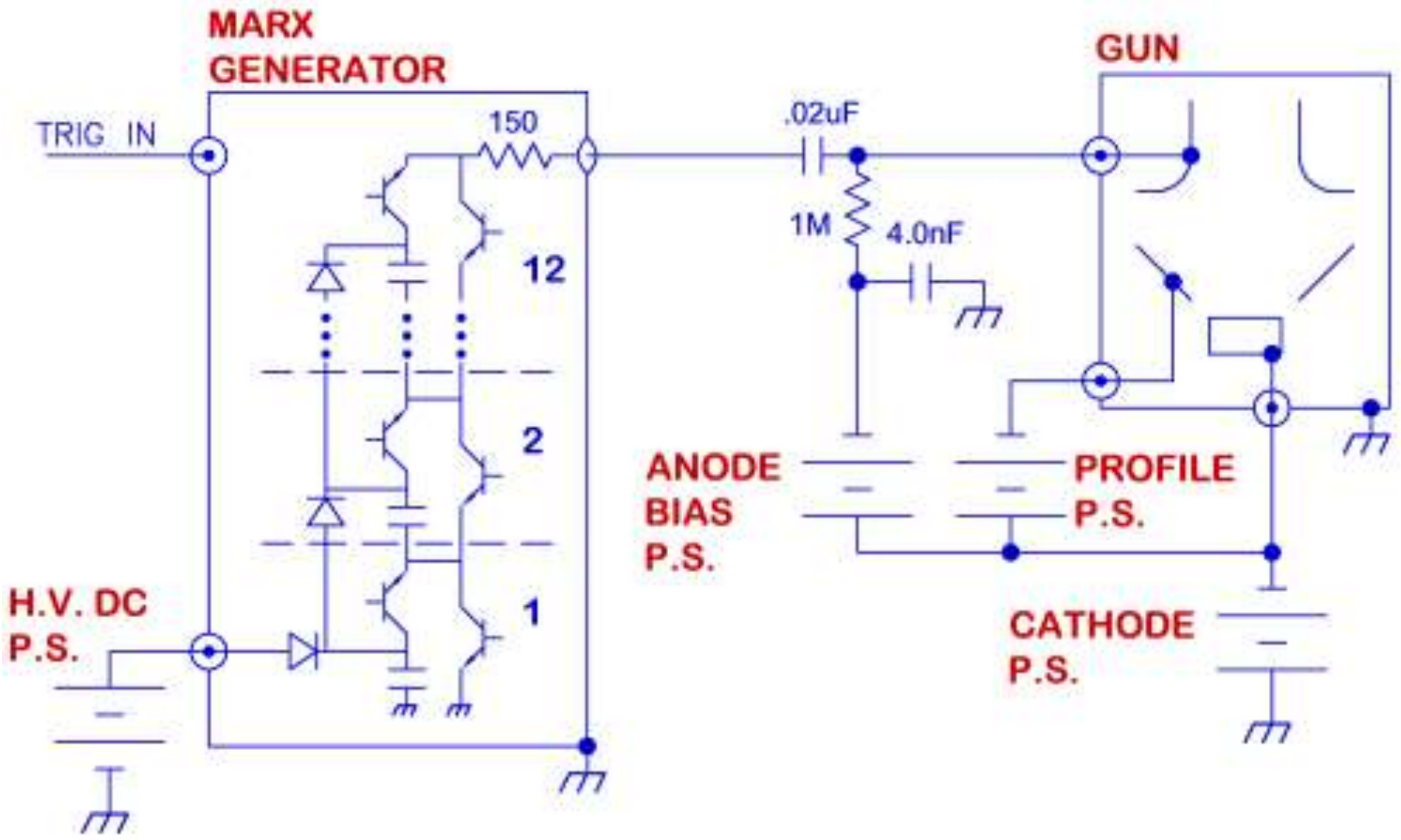

Figure 18: Electron gun driving scheme. Only the cathode, the control electrode (profiler) and the anode of the electron gun are shown.

A major design issue in the system has been the excessive switching losses in the IGBT's while running at a ~50 kHz rep rate. The device is air-cooled, and the system has run into thermal problems that have limited it to operating at less than 6 kV at the nominal rep rate. The Marx generator was designed to drive approximately 60 pF, so it must be mounted in close proximity to the gun anode terminal to avoid the extra capacitance of a long connecting cable (see Fig.18).

TEL2 is located in the Tevatron tunnel, a few yards downstream from the collider beam dump. This is arguably the worst location in the tunnel for solid-state equipment that is not radiation-hardened. The Marx generator stopped functioning in less than a week of Tevatron operation, and we found it to be Class I radioactive when we removed it from the tunnel. However, after cooling down for a few days, it started operating again. We reinstalled it behind two feet of steel shielding, and it functioned for several weeks before failing. We added four more feet of shielding, and the unit has been operating continuously for almost a year.

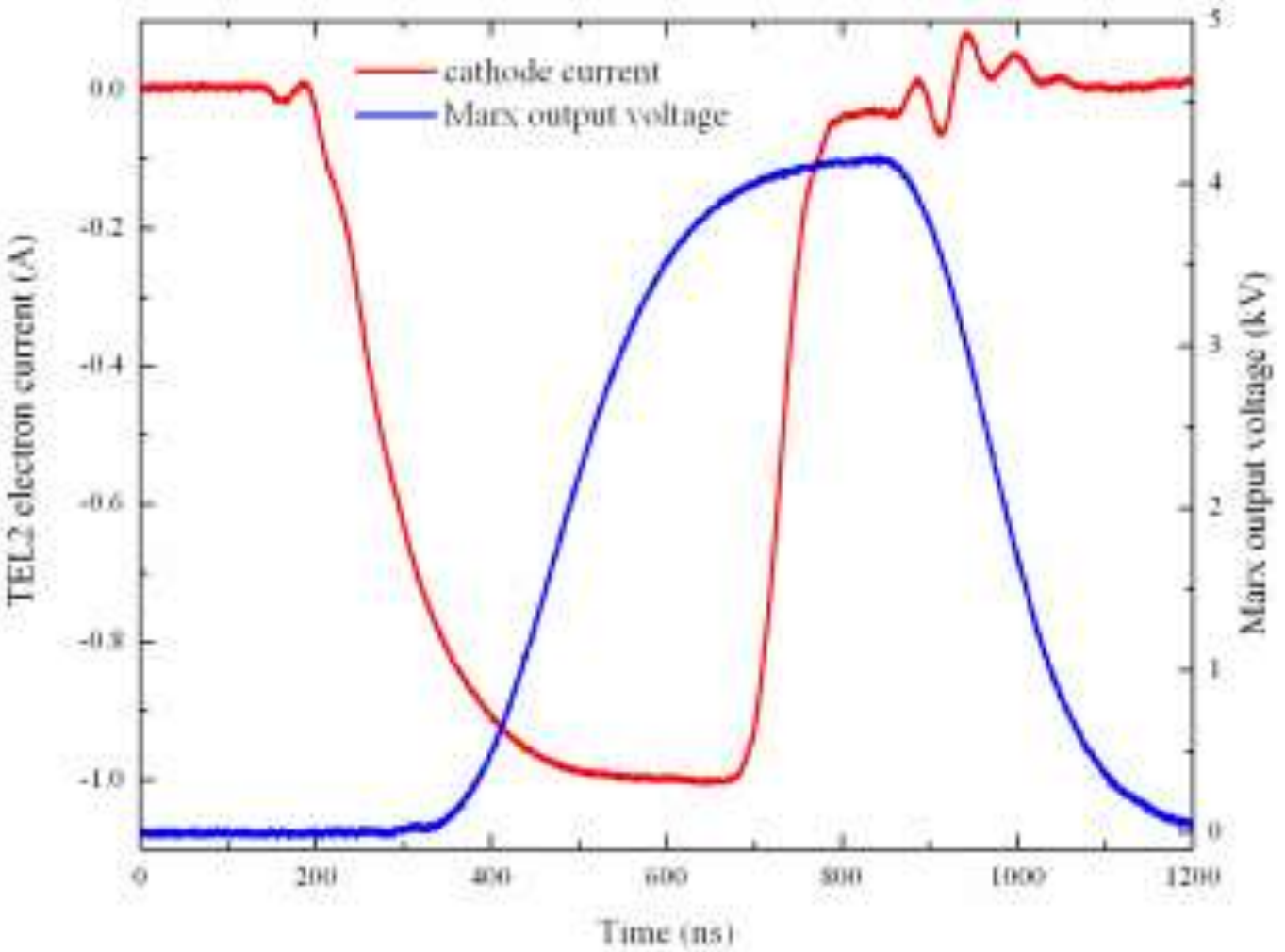

Figure 19: TEL-2 Marx generator output voltage and electron beam current.

Fig. 19 shows the output voltage of the Marx generator when gated with a 520 ns pulse. At 4.2 kV of Marx output voltage and anode bias voltage of 200 V the peak electron current is 1 A which is consistent with the SEFT gun perveance.

Currently, we are developing a second-generation Marx generator with water-cooled IGBT's with 12 different voltage levels (during one cycle, train) which will be able to work at higher voltages at the 50 kHz rep rate and a solid-state modulator based on summed pulse transformer scheme to generate similar waveforms, but at a rate of 150 kHz [20](all bunch compensation).

# 4. Beam pipes, diagnostics, other subsystems

The electron beam streaming out of the gun travels through a beam pipe until it finally reaches the collector. The beam pipe through which the electron beam travels preserves a constant inner diameter along nearly the entire length of the electron-beam path. This starts just downstream of the anode, continues around the first bend, through the main solenoid, around the second bend, and into the collector solenoid. The inner pipe diameter in the main solenoid needed to be as large as the physical aperture of the Tevatron, as the TEL was not intended to ever inhibit the performance of the Tevatron. The dotted line in Fig.20 outlines the typical aperture found around the Tevatron ring. The circumscribed circle of radius 35mm corresponds to the inner surface of the pipe through the entire length of the TELs. The TELs' vacuum components were certified according to standard

Tevatron vacuum requirements including cleaning and vacuum baking. An in-vacuum heater is installed to achieve the necessary in-situ baking temperature after the assembly and installation. Three 75 l/s ion pumps and a TSP are installed to maintain the ultra high vacuum in the system. The TEL vacuum under working conditions ranges from 8e-10 to 3e-9 Torr that is comparable with the gas pressure in the nearby sections of the Tevatron. The vacuum valves at both ends ensure that the system can be separated from the Tevatron vacuum to perform maintenance if needed.

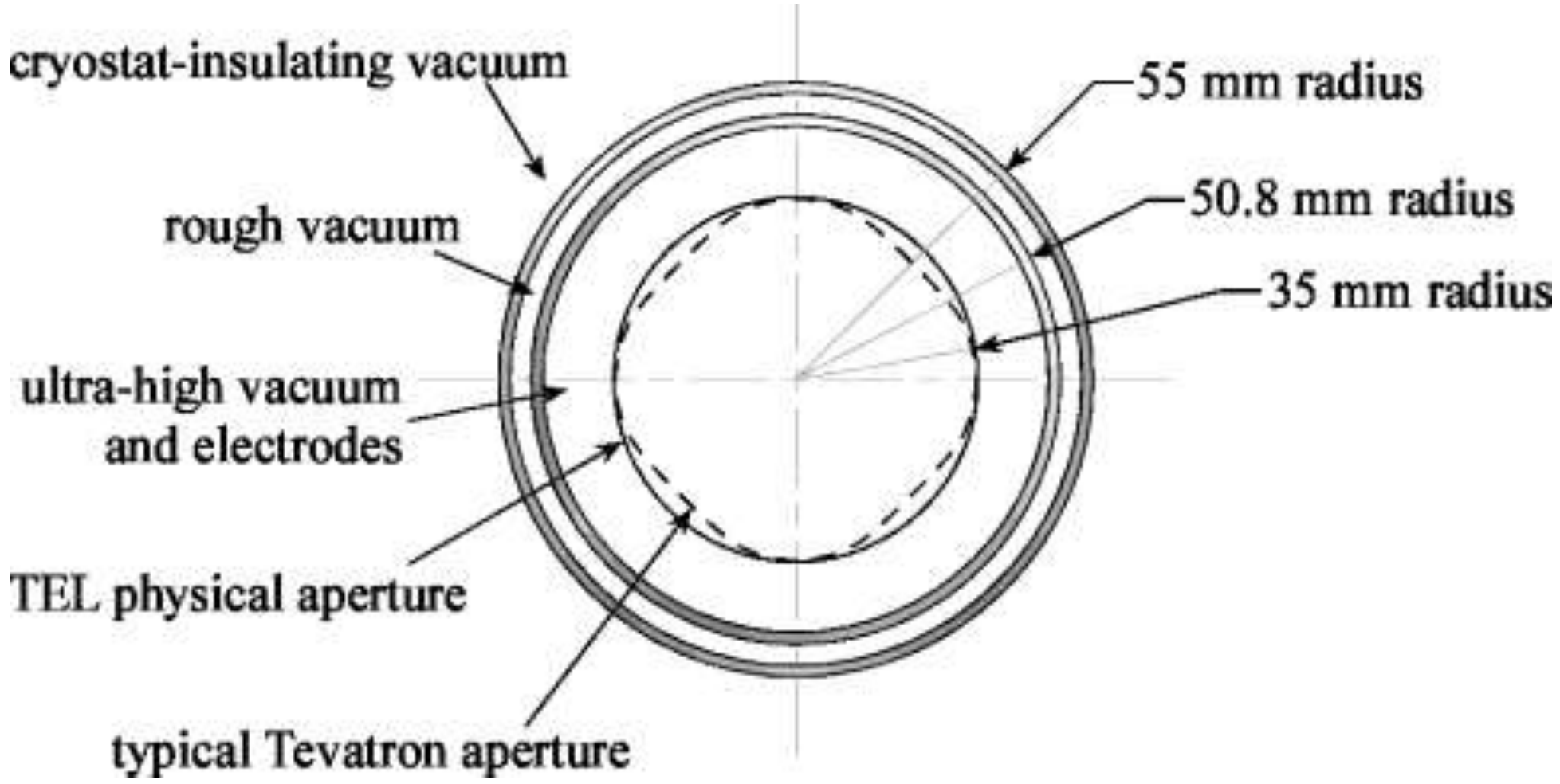

Figure 20: Cross-section of the TEL beam-pipe.

All the elements of the TEL interaction region have the same inner diameter of 70 mm as the adjacent Tevatron beam-pipes in order to minimize machine impedance. Fig.21 sketches the numerous electrodes involved in detecting and measuring the electron beam and the antiproton and proton bunches. Each of these electrodes is electrically isolated from the grounded beam pipe and is wired, through vacuum feed-throughs, to coaxial cables leading out of the Tevatron tunnel and to the support electronics. The wires shown in the center are two mechanically actuated forks. One is oriented vertically, the other horizontally, and each has a 15 mm long 0.1 mm diameter tungsten wire strung across the gap. Remotely operated motors are able to swing each fork into the middle of the beam pipe, where the electron beam is flowing. By adjusting the correctors, the beam can be swept across the fork, and the intercepted charge flows through the fork and again into cables that bring the signal out to be measured. The amount of current as a function of beam position yields data that can be converted into a profile of the current density using Abel inversion (see details and the resulting profiles in [22]). The forks are always positioned outside the beam pipe before stores begin in the Tevatron. The second TEL does not have these forks as we were more confident with the generation of the electron current profiles.

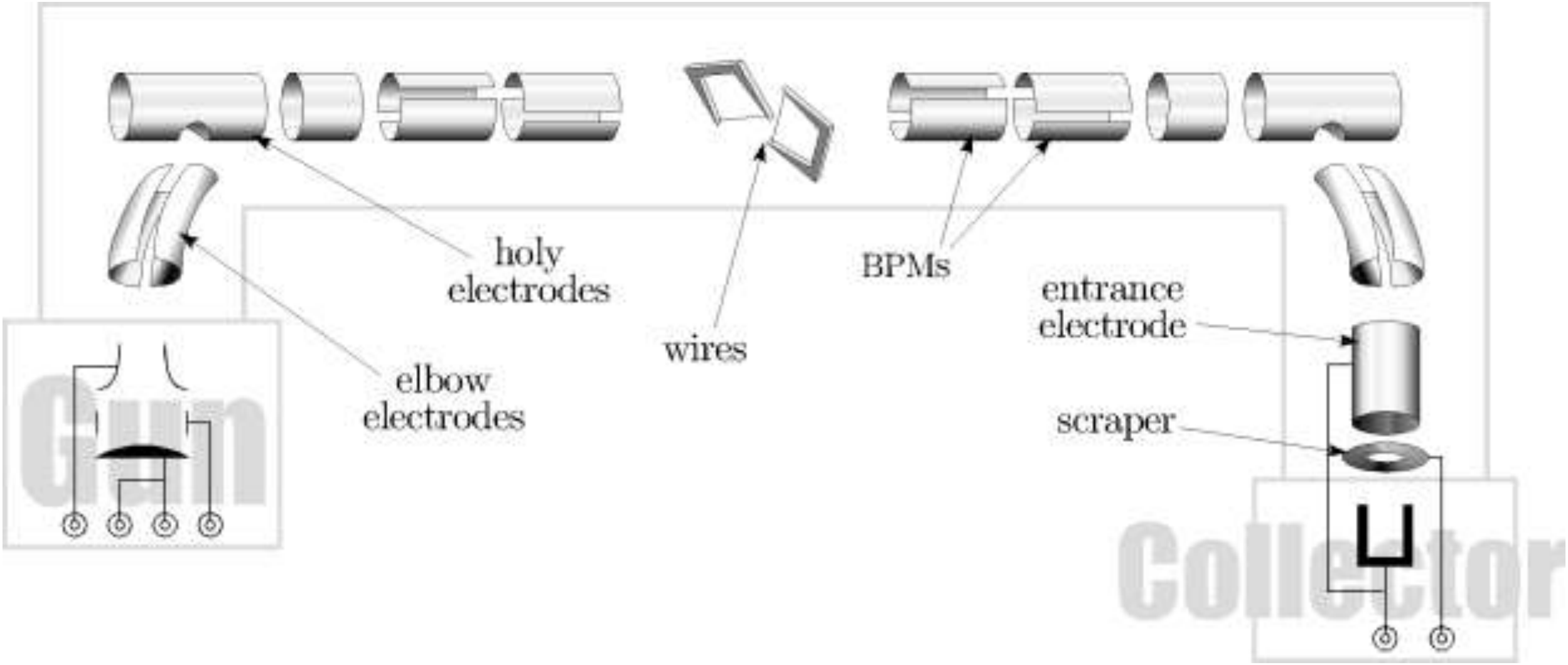

Figure 21. Sketch of the electrodes in the TEL (not to scale).

The elbow electrodes are curved imitating the path of the electron beam around each of the bends. Horizontally opposed, a high-voltage difference can be applied in order to generate a strong horizontal electric field to create a small vertical drift of the electron beam if needed, but during normal TEL operation the electrodes are grounded.

The holy electrodes are simply cylindrical electrodes that have a hole cut into one side. The electron beam passes through this hole as it enters and leaves the region of the (anti)proton orbit. These and the elbow electrodes were installed to assist with initial TEL commissioning. If the electron beam failed to pass through the solenoids and into the collector, observing which electrodes were absorbing the electron current would indicate how to correct the guiding fields. Fortunately, the electron beam had little difficulty propagating completely into the collector, and the utility of these electrodes diminished quickly.

Next to the holy electrodes in Fig.21 are cylindrical electrodes intended for clearing out ions. Ions are created by electrons bombarding residual gas molecules floating in the beam-pipe vacuum. The once-neutral molecule can easily lose electrons, turning it positively charged and attracted to the electron beam's space charge. In actuality, the influence of ions has been small enough not to induce instabilities or other problems, and these electrodes are typically grounded.

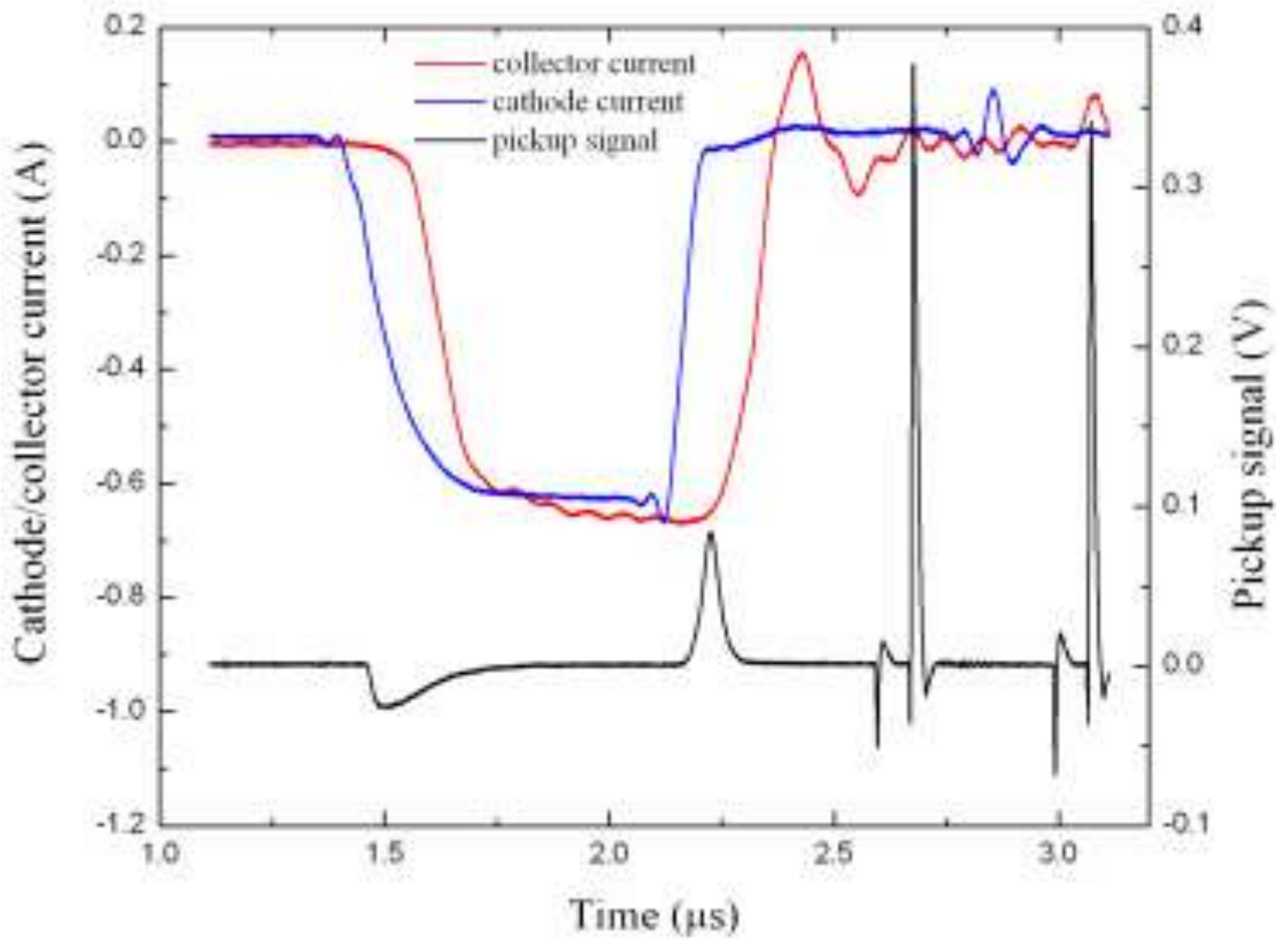

Figure 22: BPM pickup signal (black) featuring the electron pulse, two proton (positive peaks) and two antiproton bunches (negative peaks). For clarity the electron pulse is shown timed to the abort gap where no bunches are present. The scope bandwidth was intentionally set to 20MHz to reject high frequency noise and limit proton bunch signal amplitude. Cathode (blue) and collector (red) currents are measured by current transformers. The artifacts on the right hand side are caused by the instrumentation.

The TEL is equipped with four beam-position monitors (BPMs): one vertical and horizontal at the beginning and at the end of the main solenoid. The BPMs are pairs of plates that pick-up signals when any charged particle passes through them. Fig. 22 shows an example of the voltage seen on one of the BPM plates during passage of the electron pulse and few proton and antiproton bunches.

The average position of any beam charge can be calculated as:

$$x = k\frac{V_A - V_B}{V_A + V_B} \qquad (9),$$

where $x$ is the transverse distance of the beam from the center of the beam pipe, $V_{A,B}$ are the measured voltages on the two plates, and $k$=33.6 mm is a constant empirically determined from calibration measurements. To reduce the noise, each $V_i$ is calculated as the total integral of the bunch's charge profile, which in turn is the integral of the doublet current signal as charge is first pulled onto the electrode, then returned, as the bunch passes by [22].

Each 20-cm long BPM measures the beam's position in only one dimension and at only one longitudinal position. Therefore a pair of BPMs are needed on the upstream end of the main solenoid in order to record both the horizontal and vertical positions, and another pair of BPMs are needed on the downstream end. In the TELs, a fast Tektronix TDS520 oscilloscope and a LabVIEW application program operating on a dedicated PC computer process the waveform signals and compute positions constantly during stores. The design of this system includes the ability to measure the position of antiproton bunches and proton bunches with the same BPMs plates as the electron beam. In this manner, it becomes possible to confirm that the electron beam and the (anti)proton bunches are collinear within the main solenoid.

The four position detectors are sequentially connected to the oscilloscope's inputs through the Keithley RF multiplexer. The computer communicates with the oscilloscope and the multiplexer through a GPIB interface and links with Tevatron accelerator control net through Ethernet. A beam synchronous pulse generated by a standard Tevatron synchronization CAMAC module triggers the scope’s main sweep to start and digitize over either 10-20 ns of proton or antiproton bunch signal or some 1-2 μs of the electron signal. The system averages measurements over hundreds of turns in order to report positions with low error bars. The BPM signals acquired by the digital scope are processed by a LabView program which calculates positions of all three beams . Typical statistical position measurement error is about ~10-20μm peak to peak.

The BPM pickups installed in the TEL-1 are diagonal cut cylinder type which have shown an exceptional linearity. However, an unacceptably large 1-1.5 mm discrepancy has been observed in the reported positions of 10 ns short proton or antiproton bunches and ~1 μs long electron pulse [9]. The major sources of the offset are though to be the capacitance between the two plates, different stray capacitances to ground and the cross-talking between pair of electrodes which lead to significant difference of the BPM impedances for electron beam and proton beam signals, since for proton-like signal the main frequency component is about 53MHz while for electron beam the main frequency component is less than 2MHz.

Correspondingly, the following electron lens, TEL-2, has been equipped by a new type of the BPMs with four plates separated by grounded strips (to reduce plate-to-plate cross-talk). Together with new signal processing algorithm which uses 5-20 MHz band-pass filtering for both electron and proton signals convoluted with Hanning window, the frequency dependent offset has been reduced to an acceptable level of less than 0.2 mm [23, 24].

In order to reduce the noise in the system, a special attention has been placed on the cabling from the pick-up plates all the way to the oscilloscope. Fifty-ohm coaxial in vacuum cable is attached to each BPM plate, drawn through coaxial feedthroughs in the vacuum vessel, brought out of the Tevatron tunnel, and into the BPM electrical apparatus. While the outer conductor is grounded at several places along the route (such as the vacuum feedthrough and the signal switcher) reasonable preservation of the signals has been observed.

All of these cables are also shielded in 50-ohm cabling and separated from pulsed power signals, such as the anode modulator pulses. This level of caution succeeded in preventing significant contamination of low-level signals by high-power transients.

During several years of operation, a lot of effort was put into reduction of the electron beam imperfections and noises. For example, fluctuations in the current needs to be less than one percent, in order to minimize the growth of 980 GeV (anti)protons emittance. Correspondingly, cathode and anode power supplies were stabilized by filtering power line harmonics and the Fermilab specific line at 15 Hz (cycle frequency of the FNAL 8 GeV Booster synchrotron). Timing jitter of the electron pulse of more than 1 ns translates into effective electron current variation as the electron pulse usually does not have a perfect flat top that results in a significant lifetime degradation of the Tevatron bunches interacting with the electron pulse. By replacing an electron pulse function generator and a delay card we managed to reduce the jitter to less than 1ns and resolve the proton lifetime issue.

The TEL magnets have a small effect on the 980-GeV proton-beam orbit causing its distortion around the ring of about ±0.2 mm. Most of the distortion comes from transverse fields in the TEL bends.. Quenches of the main solenoid do not disturb the Tevatron beams significantly, but the electron beam is unable to propagate through the electron lenses. Hence, the interlock system turns off anode modulator power supplies in order to prevent electron beam generation. These power supplies are turned off if any corrector power supply HV power supply, or vacuum gauge detects any malfunction or anomaly. The broadband impedance of the TEL components is $|Z/n|<0.1\ \Omega$, much less than the total Tevatron impedance of $5\pm3\ \Omega$ - and correspondingly, is of no harm to the collider beams.

A very important part of the TEL operation are the diagnostics of the Tevatron bunches themselves. Monitoring of the position, intensity, losses, emittances, betatron tunes, chromaticities of high energy bunches provides extremely useful information which allows optimal tune-up of the TELs. A summary of the Tevatron beam diagnostics can be found in [25]. As the TEL electron pulse can be timed on an individual bunch, the Tevatron diagnostics instruments need to be able to work on bunch-by-bunch basis. Not all instruments can do that yet and new ones are being developed [26].

## 5. Conclusions

We have developed electron lenses - a novel type of instrument for modern accelerators. They employ space-charge forces of low energy electron beam that produce very significant and useful effects on high-energy beams (Fig.23 illustrates a tuneshift of -0.003 of the 980 GeV antiprotons in the Tevatron matching Eq.(1)). Originally designed for the purpose of the compensation of beam-beam effects in the Fermilab's Tevatron and other high-energy hadron colliders, the lenses improve the lifetime of the Tevatron proton bunches which otherwise suffered from collisions with antiproton bunches [3]. The observed proton lifetime has increased by over a factor of two at the beginning of high-energy physics collider stores, when beam brightness and luminosity are the highest and the beam-beam interaction is the strongest [4].

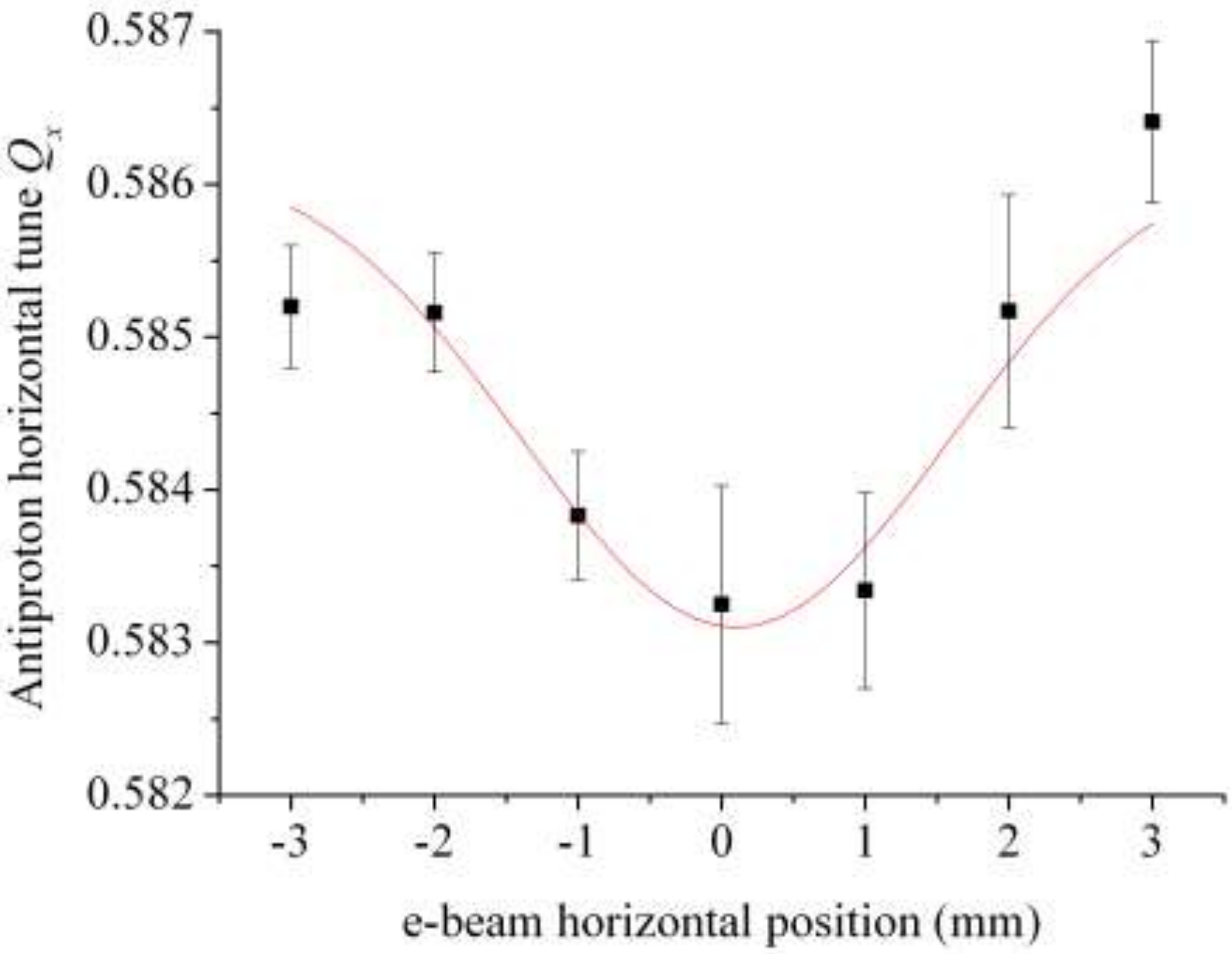

Figure 23: Horizontal betatron tune of the 980 GeV antiproton beam measured by the 1.7 GHz Schottky detector [25], plotted versus the electron beam horizontal position with respect to the antiproton orbit. "Gaussian" electron gun. Electron rms beam size 0.7mm, current of $J_e$=0.53A, cathode voltage $U_e$=7.0 kV. Electrons defocus antiprotons, thus, negative shift of the betatron tune.

Versatility of the electron lenses allows to use them for many other purposes, too. For example, TEL-1 was used for removing unwanted DC beam particles out of the Tevatron abort gaps.[5]. On several occasions, the TEL was used for elimination of improperly loaded bunches of protons and/or antiprotons and as part of tune measurement system [27]. There are proposals to use electron lenses for space-charge compensation in high intensity proton synchrotrons [28], for reduction of a tune-spread in colliding beams [29] and beam collimation in the LHC [30].

In this article we have described main elements of the Tevatron electron lenses – magnets, electron guns, collector, electron beam modulators, vacuum system and diagnostics, etc. Depending on the application of the electron lens, these elements can be modified for the purposes other than compensation of beam-beam effects.

**Acknowledgements**

We are thankful to many our colleagues and collaborators which helped in the design, fabrication and testing the two Tevatron Electron Lenses - T.Bolshakov, A.Shemyakin, S.Nagaitsev, C.Crawford, R.Hren, R.Hively, J.Featherstone, J.Fitzgerald, A.Makarov, S.McCormack, T.Andersen, F.Niell, A.Klebaner, A.Chen, A.Makarov, D.Plant, T.Johnson, J.Santucci, Y.Terechkine (FNAL), A.Sharapa, L.Arapov, T.Andreeva, P.Logatchov, B.Sukhina, B.Skarbo, V.Dudnikov, A.Larionov, Yu.Valyaev, A.Sleptsov, A.Kuzmin, and A.Aleksandrov of Budker INP (Novosibirsk, Russia), A.Ageev, A. Andriischin, A. Baluyev, I.Bogdanov, E.Kashtanov, N.Krotov, V.Pleskach, P.Shcherbakov, A.Tikhov, S.Zintchenko, V.Zubko of IHEP (Protvino, Russia), V.Efanov, P.Yarin (FID Technologies), S.Sorsher (Hi-Tech MFG, Shiller Park, IL), R.Cassel, and S.Hitchcock (Stangene Ind., CA).

TABLE I. Electron Lens and Tevatron Collider parameters.

| Parameter | Symbol | Value | Unit |
|---|---|---|---|
| | | *Tevatron Electron Lens* | |
| *e*-beam energy (oper./max) | $U_e$ | 5/10 | kV |
| Peak *e*-current (oper./max) | $J_e$ | 0.6/3 | A |
| Magnetic field in main solenoid | $B_m$ | 30.1 | kG |
| Magnetic field in gun solenoid | $B_g$ | 2.9 | kG |
| *e*-beam radius in main solenoid | $a_e$ | 2.3 | mm |
| Cathode radius | $a_c$ | 7.5 | mm |
| *e*-pulse repetition period | $T_0$ | 21 | μs |
| *e*-pulse width, "0-to-0" | $T_e$ | 0.6 | μs |
| Interaction length | $L_e$ | 2.0 | m |
| | | *Tevatron Collider* | |
| Circumference | $C$ | 6.28 | km |
| Proton(*p*)/antiproton(*a*) energy | $E$ | 980 | GeV |
| *p*- bunch intensity | $N_p$ | 270 | $10^9$ |
| *a*- bunch intensity (max.) | $N_a$ | 50-100 | $10^9$ |
| Number of bunches | $N_B$ | 36 | |
| Bunch spacing | $T_b$ | 396 | ns |
| *p*-emittance (normalized, rms) | $\varepsilon_p$ | ≈2.8 | μm |
| *a*-emittance (normalized, rms) | $\varepsilon_a$ | ≈1.4 | μm |
| Max. initial luminosity/$10^{32}$ | $L_0$ | 3.15 | $cm^{-2}s^{-1}$ |
| Beta functions at A11 TEL | $\beta_{y,x}$ | 150/68 | m |
| Beta functions at F48 TEL | $\beta_{y,x}$ | 29/104 | m |
| *p*-head-on tuneshift (per IP) | $\xi^p$ | 0.010 | |
| *a*-head-on tuneshift (per IP) | $\xi^a$ | 0.014 | |
| *p*-long-range tuneshift (max.) | $\Delta Q^p$ | 0.003 | |
| *a*-long-range tuneshift (max.) | $\Delta Q^a$ | 0.006 | |

TABLE II: Main parameters of the TEL guns

| *Parameter* | *Gun #1* | *Gun #2* | *Gun #3* | *Units* |
|---|---|---|---|---|
| Cathode diameter | 10 | 10 | 15 | mm |
| Current profile | rectangular | Gaussian | SEFT | |
| Gun perveance, max | 5.9 | 1.7 | 4.2 | $\mu A/V^{3/2}$ |
| Max. current density | 6.3 | 4.8 | 3.0 | $\mu A/V^{3/2}/cm^2$ |
| Control voltage to shut off | 2.5 | 3 | 3.5 | $U_{control}/U_{anode}$ |
| Filament power | 35-45 | 35-45 | 60-70 | W |
| *B*-field on cathode | 1 - 4 | 1 – 4 | 1 - 4 | kG |